\documentclass[11pt]{iopart}

\usepackage[utf8]{inputenc}

\usepackage{amsfonts, amssymb, latexsym}
\usepackage{graphicx}
\usepackage{rotating}
\usepackage[sort&compress,numbers]{natbib}
\usepackage{iopams}
\usepackage{xspace}

\eqnobysec

 \usepackage[usenames]{color}

\newcommand{\be}{\begin{equation}}
\newcommand{\ee}{\end{equation}}
\newcommand{\beq}{\begin{eqnarray}}
\newcommand{\eeq}{\end{eqnarray}}

\usepackage[unicode,breaklinks]{hyperref}

\begin{document}

\title[]{Testing the black hole ``no-hair'' hypothesis}


\date{\today}

\author{
Vitor~Cardoso$^{\star, \dagger, \ddagger}$,
Leonardo Gualtieri$^{\bullet}$
}
\address{$^{\star}$ CENTRA, Departamento de F\'{\i}sica, Instituto Superior T\'ecnico, Universidade de Lisboa, Avenida Rovisco Pais 1, 1049 Lisboa, Portugal}
\address{$^{\dagger}$ Perimeter Institute for Theoretical Physics, 31 Caroline Street North
Waterloo, Ontario N2L 2Y5, Canada}

\address{$^{\ddagger}$ Theoretical Physics Department, CERN, CH-1211 Gen\`eve 23, Switzerland}

\address{$^{\bullet}$ Dipartimento di Fisica, ``Sapienza'' Universit\`a di Roma \& Sezione INFN Roma 1, P.A. Moro 5, 00185, Roma, Italy}

\eads{
\mailto{vitor.cardoso@ist.utl.pt},
\mailto{Leonardo.Gualtieri@roma1.infn.it}
}

\begin{abstract}
Black holes in General Relativity are very simple objects. This property, that goes under the name of ``no-hair,'' has been refined in the last few decades and admits several versions. The simplicity of black holes makes them ideal testbeds of fundamental physics and of General Relativity itself. Here we discuss the no-hair property of black holes, how it can be measured in the electromagnetic or gravitational window, and what it can possibly tell us about our universe.
\end{abstract}



\tableofcontents
~
\clearpage
\newpage

\section{What {\it is} the no-hair hypothesis (and why should we care)?}

\subsection{The simplicity of black holes in General Relativity\label{sec:Sch}}

We celebrate this year the first direct detection of gravitational waves (GWs) and the first detection of a black hole (BH) binary,
in its last stages of coalescence~\cite{Abbott:2016blz}. In this context, it is appropriate to also honor the centenary of the Schwarzschild solution, which
describes any regular asymptotically flat, static and spherically symmetric vacuum spacetime
in General Relativity (GR). In standard Schwarzschild coordinates, the solution reads
\be
ds^2=-c^2\left(1-\frac{2GM}{c^2r}\right)dt^2+\left(1-\frac{2GM}{c^2r}\right)^{-1}dr^2+r^2d\Omega^2\,,\label{metric_Schwarzschild}
\ee
where $G$ is Newton's constant and $c$ the speed of light. The most distinctive feature of the solution above is a coordinate singularity at $r=2GM/c^2$, describing a null surface (the event horizon) which causally separates the inside and outside regions.
For physical setups where (static, spherically symmetric) matter is present, the solution should be truncated at the radius of
the object and merged smoothly with some interior solution. For pure vacuum spacetimes, it describes a BH.
In other words, any static BH in the Universe which is non-spinning and that lives in approximately empty surroundings is described
by the geometry~(\ref{metric_Schwarzschild}).

The Schwarzschild solution is fully characterized by a single parameter, the total gravitational mass $M$. In this respect, it is not dissimilar from its Newtonian counterpart: a spherically symmetric, vacuum, static solution of Newton's gravity is also described by only a mass parameter. It turns out to be difficult to construct BH solutions described by more parameters. For example,
let's try to ``anchor'' a weak, {\it static} massless spin-$s$ field onto the Schwarzschild solution. For the sake of illustration we focus on minimally coupled scalars, vector fields described by Maxwell's theory and gravitational fluctuations within vacuum GR~\cite{Berti:2009kk}. At a linearized level (i.e., keeping the background geometry fixed), these fields 
can be expanded in scalar, vector or tensor harmonic functions, parametrized by an integer number $l=0,1,2,...$, which carry information on the angular dependence of the field. 
These fields are all described by the equation~\cite{Berti:2009kk}
\be
\left(\left(1-\frac{2GM}{c^2r}\right) \Psi'\right)'-\left(\frac{l(l+1)}{r^2}+\frac{2GM(1-s^2)}{c^2r^3}\right)\Psi=0\,,
\ee
where $\Psi$ denotes the field amplitude,  $s=0,1,2$ for scalars, vectors and tensors, respectively, and primes stand for radial derivatives. One can now multiply the above equation by the complex conjugate $\Psi^*$, integrate from the horizon to infinity, and look for regular solutions of the above equation. We get, upon performing an integration by parts and dropping a boundary term (regular solutions evaluate it to zero),
\be
-\int_{\frac{2GM}{c^2}}^{+\infty}dr\left(1-\frac{2GM}{c^2r}\right)|\Psi'|^2+\frac{l(l+1)}{r^3}\left(r+\frac{2GM(1-s^2)}{c^2\,l(l+1)}\right)|\Psi|^2=0\,.
\ee
For any $l\geq s$ the integrand is negative-definite outside the horizon, and the only solution to the above equation is the trivial one, $\Psi=0$.
For vector fields, there is a nontrivial solution $\Psi={\rm const}$ for $l=0$, describing a weakly charged
BH. For tensors, the only non-trivial solutions (for which the integrand is not positive-definite) have $l=0, 1$ and correspond to a slight change of mass and addition of small amount of 
angular momentum~\cite{Zerilli:1971wd}.  This simple exercise then leads one to the following conclusions:

\noindent (i) It is impossible to ``anchor'' non-interacting scalars (and turns out, also fermions) onto a Schwarzschild BH.


\noindent (ii) The Schwarzschild solution allows, in principle, for a generalization that includes (only) electric charge and rotation.

Note that the conclusions above are drawn in the context of GR; other theories would lead to different equations of motion that could (and do, sometimes) lead to other types of solutions.
Thus, non-rotating BHs do not exhibit any ``protuberance.'' John Wheeler and others summarized these results with the expression ``black holes have no hair'', where ``hair'' is a measure of the complexity of the gravitational field of the geometry.
We will soon see that, within GR, BHs can rotate and be electromagnetically charged. Thus in fact BHs have a finite number of hairs, although the term ``no-hair'' is still loosely applied to these solutions as well. In general, as we show below, if one characterizes the geometry through its multipole moments, BHs have only a finite and small number of independent quantities that suffice to completely describe its multipolar structure. 
\subsection{The uniqueness and no-hair theorem(s)\label{subsec:theorems}}

In fact, a charged and rotating BH solution was discovered many decades after Schwarzschild's work,
and is now known as the Kerr-Newman BH, carrying mass, electric charge and angular momentum~\cite{Kerr:1963ud,Newman:1965my}.
In addition, in a series of uniqueness theorems 
(see~\cite{Bekenstein:1996pn,Carter:1997im,Chrusciel:2012jk,Robinson:1975bv,Robinson}
for reviews), it was established,

\vskip 5mm
\noindent {{\it Theorem 1}}: an isolated, stationary and regular BH in Einstein-Maxwell
theory is described by the Kerr-Newman family. 

\vskip 5mm

In other words, the structure of asymptotically flat, stationary BHs of Einstein-Maxwell theory is completely determined by its {\it global charges} defined at infinity, in particular its mass, angular momentum and electric charge. Astrophysical BHs are thought to be neutral to a very good
approximation, because of quantum discharge effects, electron-positron pair
production and charge neutralization by astrophysical plasmas~\cite{Barausse:2014tra}~\footnote{These mechanisms are suppressed in theories of minicharged dark matter, when the charged particles have a charge-to-mass ratio much smaller than that of the electron, and for which BHs may be charged
under a hidden $U(1)$ symmetry~\cite{Cardoso:2016olt}.}. Because of this, we will focus almost entirely on electrically neutral geometries. The neutral version of the Kerr-Newman solution
is simply described by the two-parameter Kerr metric~\cite{Kerr:1963ud}, which in standard
Boyer-Lindquist coordinates reads
\begin{eqnarray}
ds^2&=&-c^2\left(1-\frac{2GMr}{c^2\Sigma}\right)dt^2+\frac{\Sigma}{\Delta}dr^2+\Sigma d\theta^2 -\frac{4GJ r\sin^2\theta}{c^2\Sigma}dtd\phi\nonumber\\ 
&+&\left[r^2+\frac{J^2}{M^2c^2}+\frac{2GJ^2 r\sin^2\theta}{Mc^4\Sigma}\right]\sin^2\theta d\phi^2\,, \label{metricKerr}
\end{eqnarray}
where $\Sigma\equiv r^2+\frac{J^2}{M^2c^2}\cos^2\theta,\,\Delta=r^2-\frac{2GMr}{c^2}+\frac{J^2}{M^2c^2}$.

Note that Theorem 1 does {\it not} require axi-symmetry, it being a consequence of stationarity. 
Note also that Theorem 1 as well as all results discussed in the text, makes further analiticity assumptions, which we do not discuss here (but see Refs.~\cite{Ionescu:2011wx,Ionescu:2015dna}). Due to these assumptions (which are required to infer axi-symmetry from stationarity), there are also those who advocate that the no-hair theorems are less general than commonly stated (see~\cite{Dafermos:2008en} and references therein).

From now on, we will use geometric units with $G=c=1$, and express the Kerr solution as above with
%
%
%
$\Sigma\equiv r^2+a^2\cos^2\theta,\,\Delta=r^2-2Mr+a^2$. This metric describes the gravitational field of a spinning BH of mass
$M$ and angular momentum $J=a M$. The event horizon is located at $r_+=M+\sqrt{M^2-a^2}$, and the BH spin is bounded from above by $|a|\leq M$. 

This uniqueness result is also an example of a no-hair theorem, and the horizon plays a crucial role
in exclusion of ``hair,'' as it prevents information traveling at finite speeds (other then that conveyed by conserved charges)
to cross the horizon. This expectation, originally conjectured by Wheeler, was put on firmer ground in a series of works.
For example, Hartle~\cite{Hartle:1971qq} and Teitelboim~\cite{Teitelboim:1972ps,Teitelboim:1972pk,Teitelboim:1972qx}
showed that it is impossible to measure baryon or lepton numbers of BHs. In flat space there is a long-range $1/r^5$ potential between two collections of matter arising from the exchange of neutrino pairs between them. 
But once one of the neutrinos crosses the horizon, a weak-interaction force of this kind ceases to exist. A Kerr BH has no exterior neutrino field with classical effects.

In fact, two strong results emerged regarding BHs in the presence of fundamental fields. The first concerns the existence of stationary BHs surrounded by stationary matter, in the form of fundamental fields without self-interactions\footnote{see Section~\ref{sec:hairy} for counter-examples involving other forms of matter like anisotropic fluids.},

\vskip 5mm

\noindent {\it Theorem 2}: an isolated, stationary and regular BH in the Einstein-Klein-Gordon or Einstein-Proca\footnote{The Einstein Proca theory describes a theory with a massive photon, see e.g., Ref.~\cite{Witek:2012tr}.} theory with a {\it time-independent boson} is described by the Kerr family~\cite{Bekenstein:1972ky,Hawking:1972qk,Sotiriou:2011dz,Graham:2014ina}.
\vskip 5mm
The fields can be also massless and complex, but the theorem requires them to inherit the spacetime symmetries, i.e., no time-dependence.
Physically, time-dependent fields will scatter to infinity and/or enter the horizon and therefore break the assumed stationarity. 
A non-trivial time-dependence can be excluded for {\it real} scalars, 

\vskip 10mm

\noindent {\it Theorem 3}: an isolated, stationary and regular BH in the Einstein-Klein-Gordon 
theory with one {\it real} scalar is described by the Kerr family~\cite{Bekenstein:1972ky,Hawking:1972qk,Sotiriou:2011dz,Graham:2014ina}. 
\vskip 5mm

Real scalars with a nontrivial time dependence will give rise to a nontrivial quadrupole moment, therefore emitting GWs, again breaking the stationarity assumption.
Notice that complex scalars are able to avoid this result, by producing a time-independent stress-energy tensor. At the same time, they are also able to avoid being absorbed at the horizon
through a superradiant mechanism. We will see in Section~\ref{subsec:sm} that in fact complex fields are able to produce hairy BHs.

We should mention that there are also some ``no-hair'' results for dynamical BH systems, and that these may even include evolving BH binaries. We refer the reader to Ref.~\cite{Berti:2013gfa} for a discussion of the general situation, but we will not discuss these examples any further here.

In summary, the ``no-hair'' theorem(s) are a set of proofs that -- under some conditions -- Kerr-Newman is the only possible asymptotically flat and regular solution
of the field equations in the presence of fundamental fields (see Refs.~\cite{Heusler:1996ft,Chrusciel:2012jk,Berti:2015itd,Herdeiro:2015waa,Volkov:2016ehx} for further details). 

\vskip 5mm
\subsection{The ``no-hair'' or ``Kerr'' hypothesis}

However, there is no proof that Kerr-Newman is the most generic solution of the field equations (see below in Section~\ref{sec:hairy} for the reasons why such proof cannot exist). The ``no-hair'' or ``Kerr'' {\it hypothesis} states simply that the Kerr geometry -- which depends on only {\it two parameters} -- describes {\it any} BH in the universe (with the exception of those involved in highly-dynamical phenomena). This is no modest proposal!

One thus is led to two questions:

\noindent {\bf a.} If one sets up some arbitrary initial conditions how is the final, Kerr state approached?

\noindent {\bf b.} {\it Is} Kerr inevitably the final state?

\noindent {\bf c.} How can one test for a. or b.?

We will try to answer these questions below. The no-hair hypothesis is pivotal to 
interpret observations of massive astrophysical bodies. 
The only other compact object that is agreed to populate the universe are neutron stars.
But neutron stars in GR cannot be more massive than $\sim 3M_\odot$~\cite{Rhoades:1974fn}, as not even degeneracy pressure can be sustained for stars more massive than this limit. Thus, within the framework of GR with a standard matter sector, and of the no-hair hypothesis, any compact object~\footnote{By compact we mean compactnesses $2GM/c^2R\gtrsim 10^{-3}-10^{-2}$.} 
with mass larger than $\sim 3M_\odot$ is a Kerr BH. Conversely, any observation of a compact object with mass larger than $\sim 3M_\odot$ and with metric different from the Kerr geometry would inevitably signal a departure from standard physics (either in the gravitational or in the matter sector). Therefore tests of strong-field gravity targeting BH systems aim at verifying the ``Kerr hypothesis'' in various ways.

\section{The dynamics of hair-loss for massless fields\label{sec:hair_loss}}
%
\begin{figure*}[ht]
\begin{center}
\begin{tabular}{cc}
\includegraphics[width=0.50\textwidth]{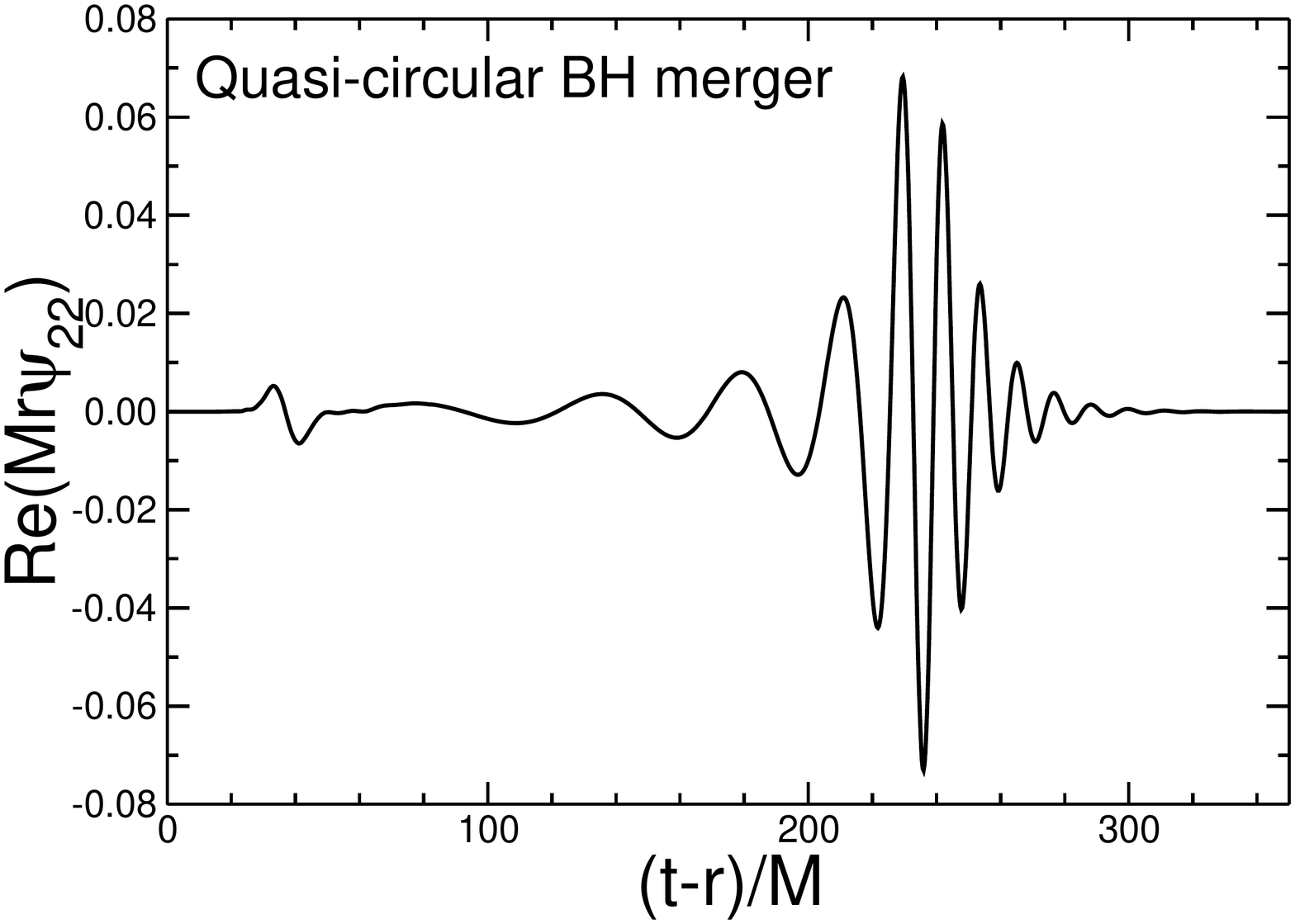}&
\includegraphics[width=0.50\textwidth]{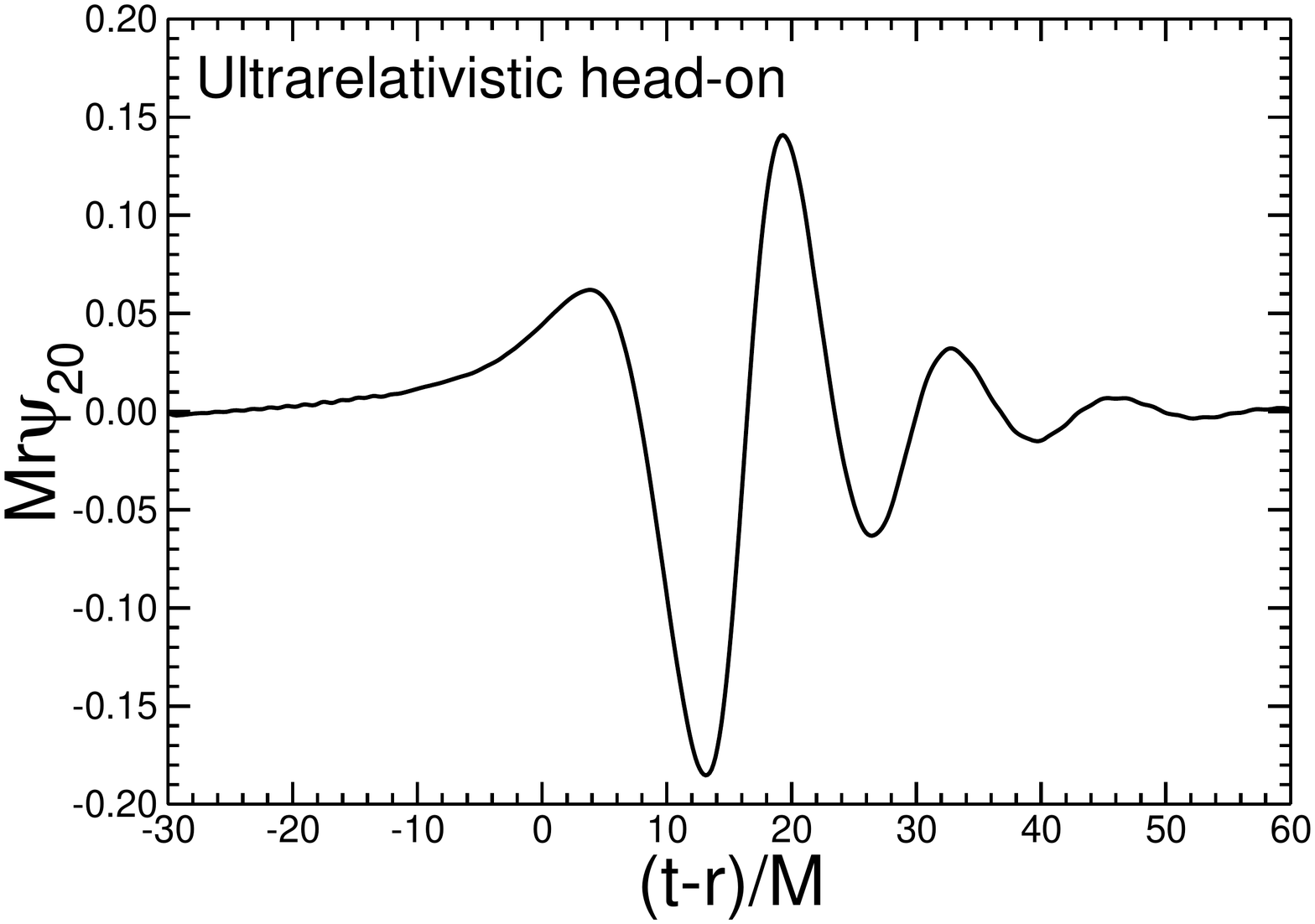}\\
\includegraphics[width=0.50\textwidth]{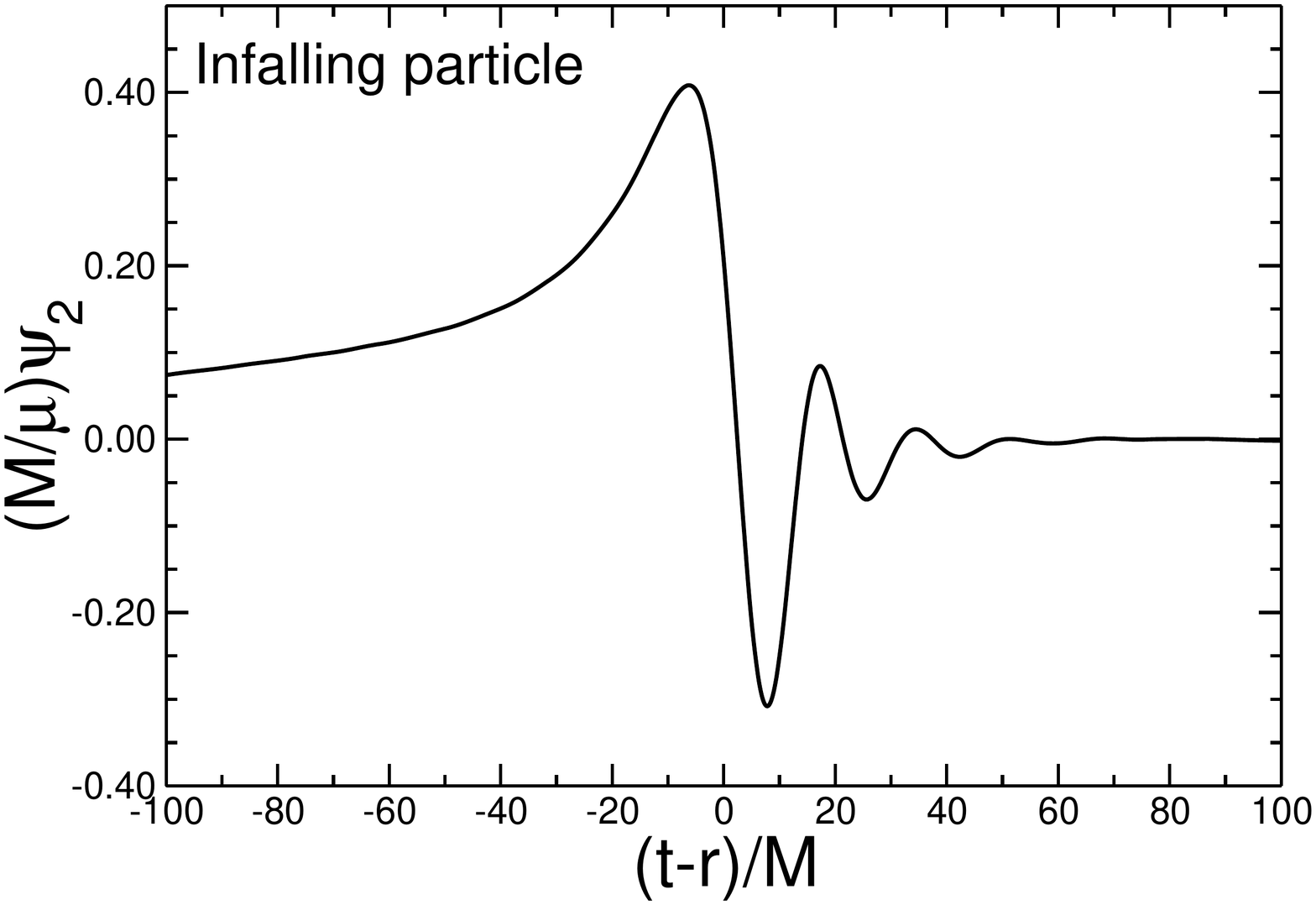}&
\includegraphics[width=0.50\textwidth]{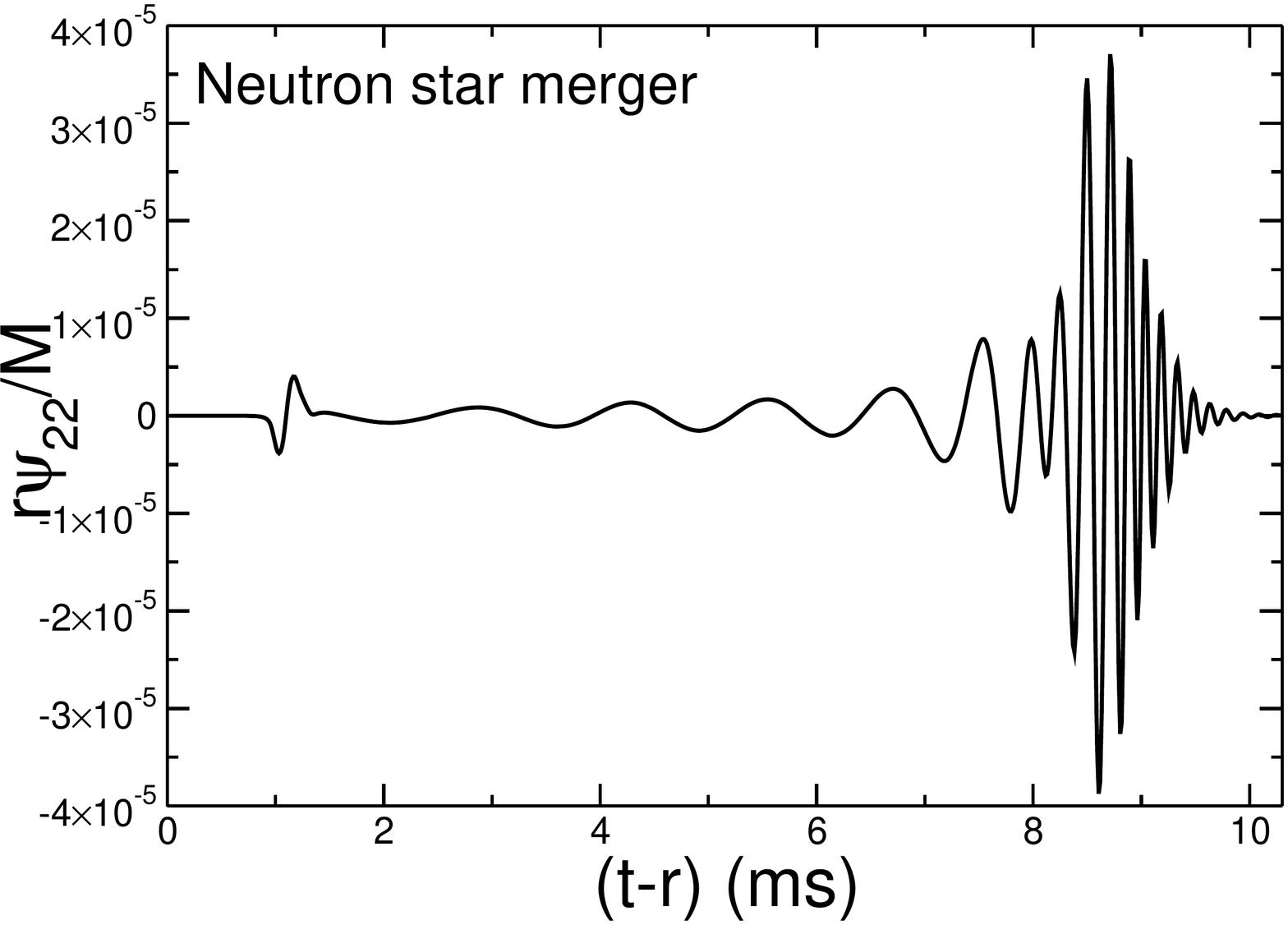}
\end{tabular}
\end{center}
\caption{Four different physical processes leading to substantial quasinormal
  ringing. In all of them, quasinormal ringing is clearly visible. 
	The {\bf upper-left} panel (adapted from Ref.~\cite{Berti:2007fi}) 
	is the signal from two equal-mass BHs initially on quasi-circular orbits, inspiralling towards
each other due to the energy loss induced by GW emission,
merging and forming a single final BH. The {\bf upper-right}
panel shows gravitational waveforms from numerical simulations of two equal-mass BHs, 
colliding head-on with $v/c=0.94$ in the center-of-mass frame: as the center-of-mass energy grows (i.e., as the
speed of the colliding BHs tends to the speed of light) the waveform is more
and more strongly ringdown-dominated \cite{Sperhake:2008ga}. The {\bf bottom-left}
panel shows the gravitational waveform (or more precisely, the dominant, $l=2$
multipole of the Zerilli function) produced by a test particle of mass $\mu$
falling from rest into a Schwarzschild BH \cite{Davis:1971gg}: the shape of
the initial precursor depends on the details of the infall, but the subsequent
burst of radiation and the final ringdown are universal features. The
{\bf bottom-right} panel (reproduced from Ref.~\cite{Baiotti:2008ra}) shows GWs emitted by two massive neutron stars with a polytropic equation of
state, inspiralling and eventually collapsing to form a single BH.
With the exception of the infalling-particle case (where $M$ is the BH mass, $\mu$ the particle's mass and $\psi_2$ the
Zerilli wavefunction), $\psi_{22}$ is the $l=m=2$ multipolar component of
the Weyl scalar $\Psi_4$, $M$ denotes the total mass of the system and $r$ the extraction radius (see
e.g. Ref.~\cite{Berti:2007fi}). Taken from Ref.~\cite{Berti:2009kk}.} \label{fig:qnmexcitation}
\end{figure*}

The first studies concerning the dynamics of BH spacetimes were mostly focused on Kerr BH backgrounds, weakly disturbed by {\it massless, minimally coupled and non-interacting} fields (see Section~\ref{sec:hairy} below for the full picture). They included the scattering of Gaussian wavepackets~\cite{vish} and the infall of a point-like particle~\cite{Davis:1971gg}. These attempts showed that the GW (or scalar or electromagnetic) 
signal from a perturbed BH can in general be divided in
three parts~\cite{Berti:2009kk} (see Fig.~\ref{fig:qnmexcitation}). The signal starts with a prompt response at early times, which depends on the initial conditions and corresponds to direct propagation of the wave from source to observer.

After the prompt response, and as the source crosses the BH light ring (i.e, the unstable circular null geodesic), it excites
the oscillation modes of the BH~\footnote{These modes can, alternatively, be thought of as a slow leakage of waves trapped in the circular null geodesic~\cite{Cardoso:2008bp}.}. The vibration modes of BHs are called quasinormal modes (QNMs) and
consist on a superposition of exponentially damped sinusoids,
\be
\Psi\sim\sum_{lmn} A_{lmn}e^{i\omega_{lmn} t+\phi_{lmn}}e^{-t/\tau_{lmn}}=\sum_{lmn} A_{lmn}e^{2\pi i f_{lmn} t+\phi_{lmn}}e^{-t/\tau_{lmn}}\,, \label{qnm_exp}
\ee
with characteristic QNM frequencies $(\omega_{lmn},1/\tau_{lmn})$ that depend only on the BH mass and spin, because the ``progenitor''
is fully characterized by these two parameters.
These frequencies are tabulated and publicly available~\cite{rdweb,Berti:2005ys,Berti:2009kk}.
The modes depend on the integers $l,\,m$ labeling the angular dependence, where $l=2,3,...$ and $m=0,\,\pm1,\,\pm2...\pm l$.
The overtone index $n=0,1,...$~\cite{Berti:2005ys,Berti:2009kk}.

For a Schwarzschild BH, the quadrupolar fundamental frequencies read
\beq
f_{220}&=&f_{200}=1.207\times 10^{-2}\left(\frac{10^6M_{\odot}}{M}\right)\,{\rm Hz}\,,\\
\tau_{220}&=&\tau_{200}=55.37\left(\frac{M}{10^6M_{\odot}}\right)\,{\rm sec}\,.\label{eq:QNMs_a0}
\eeq
\begin{figure*}[ht]
\begin{center}
\begin{tabular}{cc}
\includegraphics[width=0.50\textwidth]{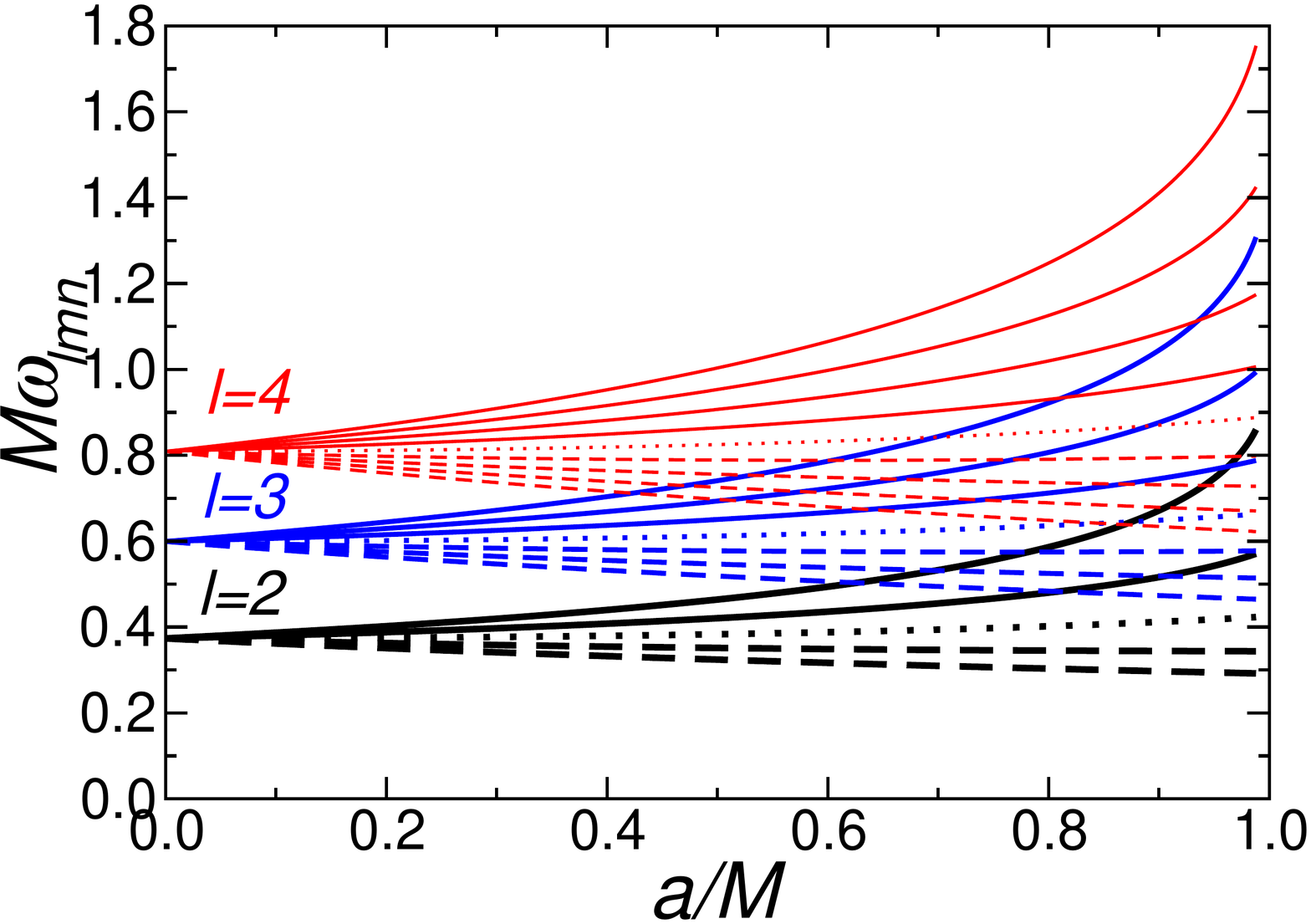}&
\includegraphics[width=0.50\textwidth]{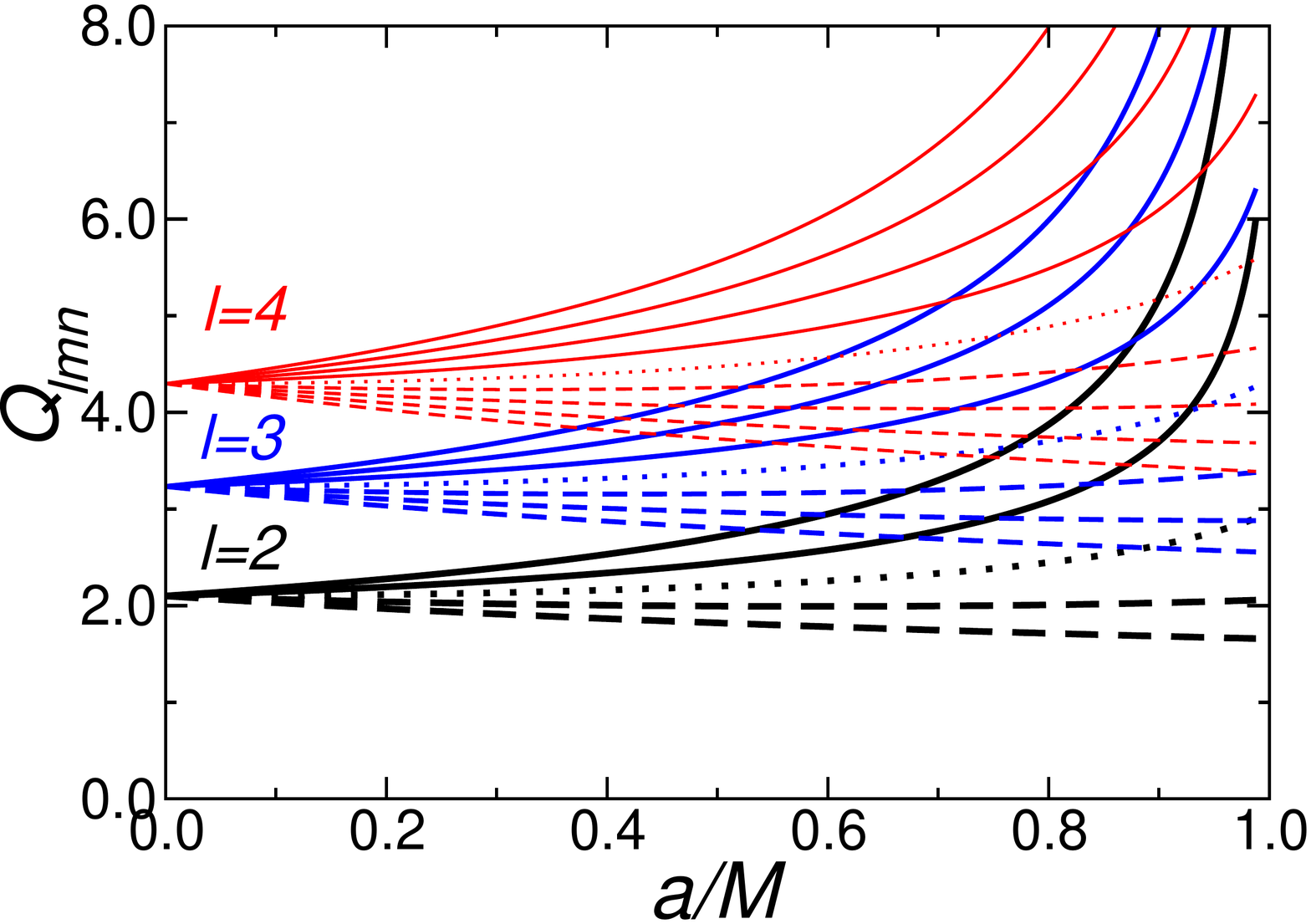}
\end{tabular}
\caption{Frequencies and quality factors for the fundamental modes with
  $l=2,~3,~4$ and different values of $m$. Solid lines refer to $m = l, .., 1$
  (from top to bottom), the dotted line to m = 0, and dashed lines refer to $m
  = -1, ..,-l$ (from top to bottom). Quality factors for the higher overtones
  are lower than the ones we display here. Taken from Ref.~\cite{Berti:2009kk}.}
\label{fig:fQKerr}
\end{center}
\end{figure*}
It is sometimes more convenient to work instead with the quality factor $Q_{lmn}=\omega_{lmn}\tau/2$,
a measure of how many ringdown cycles are contained in the signal. The frequencies and quality factor of Kerr BHs are shown in Fig.~\ref{fig:fQKerr}, as function of the dimensionless angular momentum $j\equiv a/M$. Highly spinning Kerr BHs are good resonators, the quality factor becoming very large.

In Refs.~\cite{rdweb,Berti:2005ys,Berti:2009kk} the frequencies and quality factors
of the first three overtones (for $l=2,~3,~4$ and all values of $m$) were
fitted by functions of the form
\beq
2\pi M f_{lmn}&=&f_1+f_2(1-j)^{f_3}\,,\label{ffit}\\
Q_{lmn}&=&q_1+q_2(1-j)^{q_3}\,.\label{Qfit}
\eeq
Here the constants $f_i$ and $q_i$ depend on $(l\,,m\,,n)$ (see Tables VIII-X
in \cite{Berti:2005ys}), and the fits are accurate to better than $4\%$ for $j\in [0,0.99]$.

\begin{figure}
\begin{center}
\begin{tabular}{cc}
\includegraphics[width=0.5\textwidth]{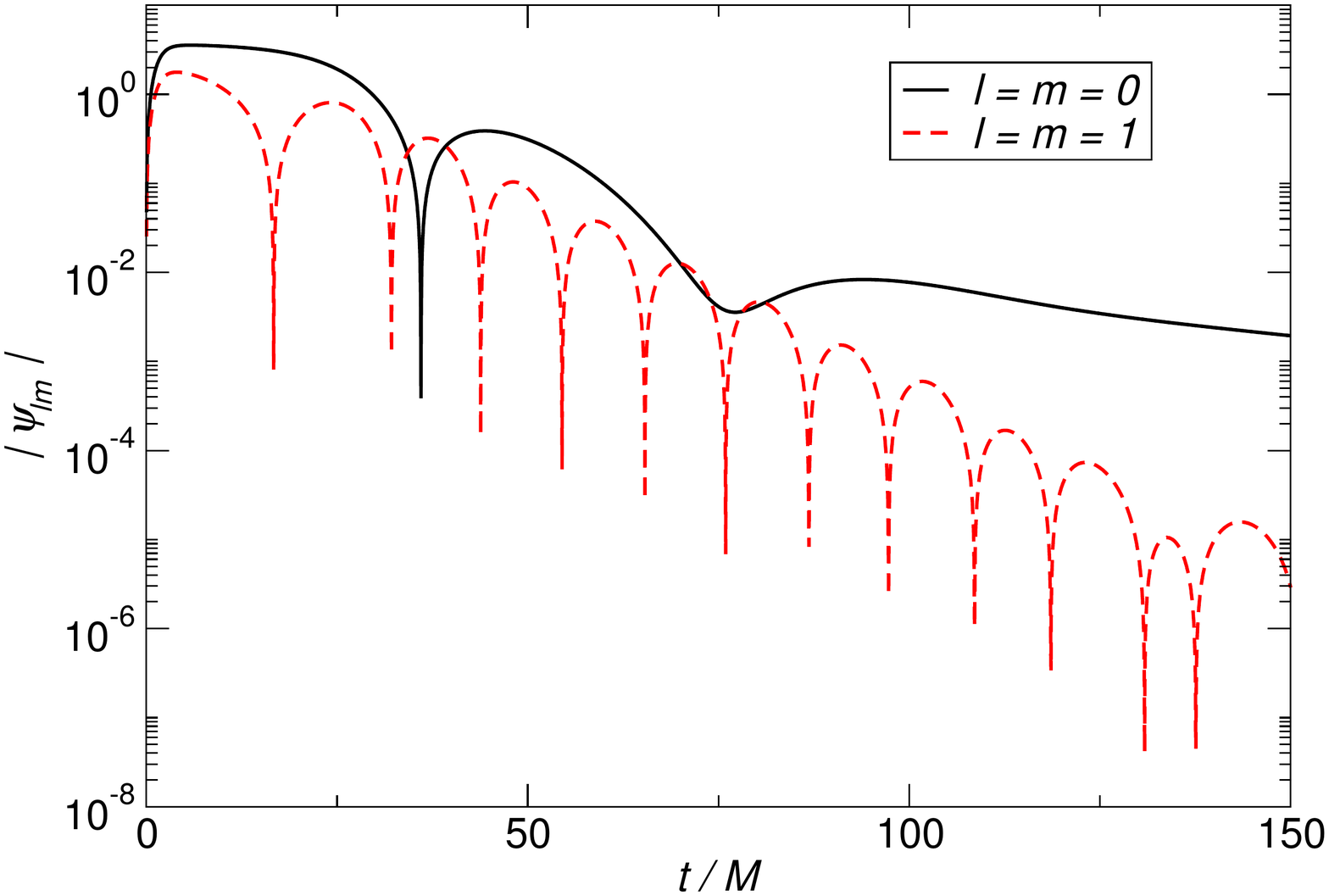} &
\includegraphics[width=0.5\textwidth]{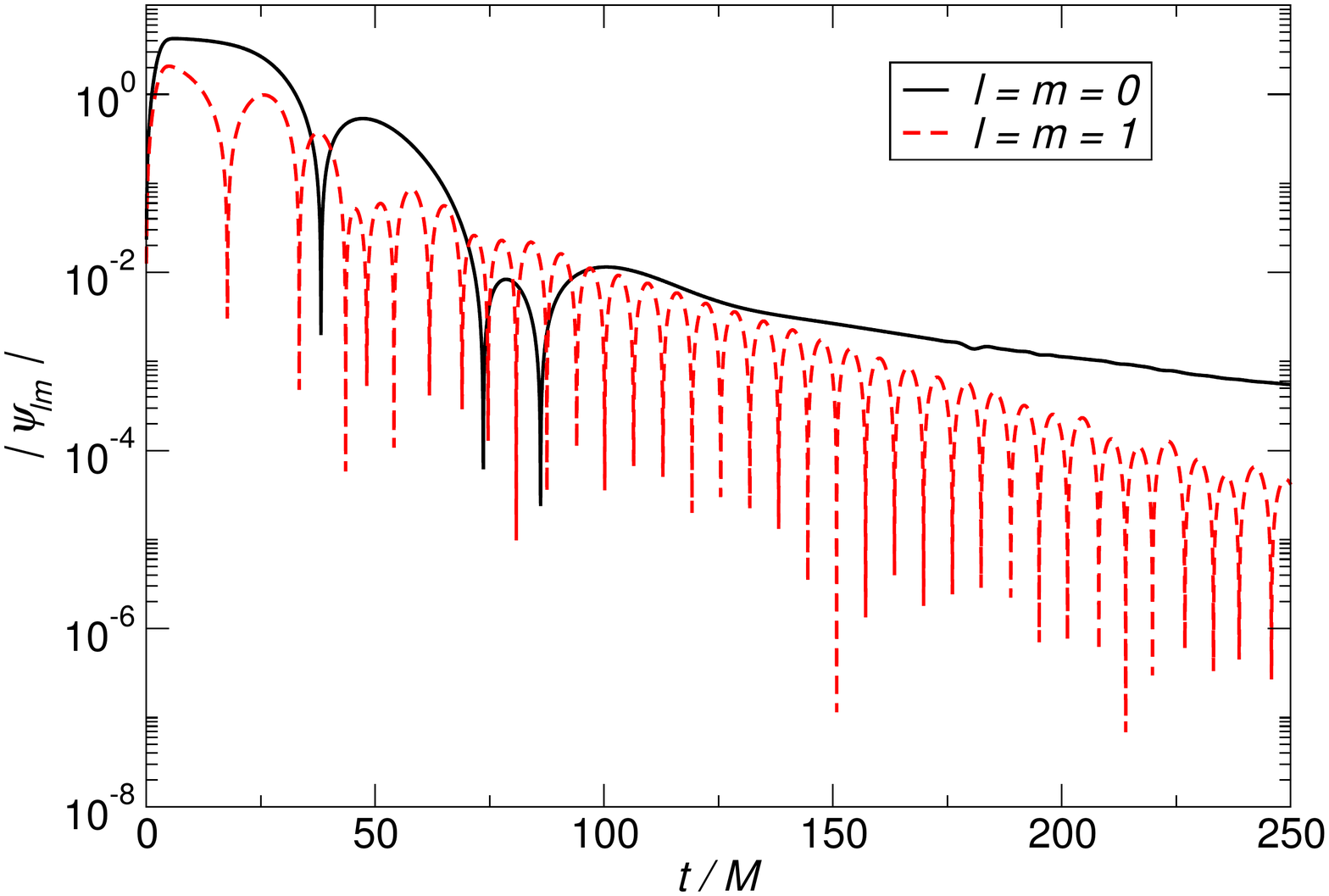} 
\end{tabular}
\end{center}
\caption{\label{fig:MFScaMLwaveform} 
Evolution of a Gaussian profile of a {\it massless} scalar field with width
$w=2~M$ centered at $r_0=12~M$ around a Schwarzschild (left panel)
and a Kerr BH with $a/M = 0.99$ (right panel).
We depict the $l=m=0$ (solid black line) and $l=m=1$ (red dashed line) multipoles.
The multipolar components of the field were extracted
at $r_{\rm ex}=10$. The waveform displays an early transient
followed by an exponentially decaying sinusoid (QNM ringdown) and a power-law tail
at late times. The late-time power-law tail has the form $t^{p}$ for the monopole,
with $p=-3.08$ for $a/M=0$ and $p=-3.07$ for $a/M=0.99$ in good agreement
with the prediction $p=-3$ obtained from the low-frequency expansion
of the wave equation (\ref{tail_massless}). From Ref.~\cite{Witek:2012tr}.
}
\end{figure}
Finally, at very late times, a power-law falloff of the field, not visible in Fig.~\ref{fig:qnmexcitation}, ensues~\cite{Price:1971fb,Ching:1995tj},
\be
\Psi\sim t^{-(2l+3)}\,.\label{tail_massless}
\ee
This decay is clearly visible in evolutions of Gaussian wavepackets, as shown in Fig.~\ref{fig:MFScaMLwaveform}.

\section{The death of the no-hair hypothesis
\label{sec:hairy}}

Although very attractive, the no-hair hypothesis is dangerously close to either being 
(i) useless, if it requires vacuum or truly stationary spacetimes, in which case there is probably no BH in our universe satisfying these assumptions or (ii) contradicted by every-day observations
of BHs surrounded by {\it quasi-stationary} matter, such as accretion disks or orbiting stars,
since these systems are not described by the Kerr solution.

In other words, observations of BHs are bound to happen in ``dirty'' environments, not in vacuum. 
For massive BHs, it can be argued -- and make the statement precise -- that accretion disks or orbiting stars can be disentangled from the BH background, by treating them as a small fluctuation. In this sense, tests of the no-hair hypothesis would still be possible,
if the full spacetime deviates only ``slightly'' from that of a Kerr BH, and the ``slightness'' is controlled by the amount of matter that we see around the BH.

We will now argue, and show, that even at a fundamental level the ``no-hair'' hypothesis has been falsified several times in the past: there exist stationary solutions of the field equations which represent BH spacetimes that are not Kerr and that cannot be mapped to Kerr with the help of some small, observationally controlled parameter.

\subsection{Anisotropic fluid hair\label{subsec:fluid}}

One intuitively expects that stationary -- or at least very long-lived -- hairy
BHs are possible if one encloses a small spherical BH at the center of a large spherical ``wall'', with pressure just enough
to keep the wall static. In other words, we do not expect anything drastic to occur if a, say, $1\,{\rm Kg}$ or $10^{-24}\,{\rm mm}$ BH is placed inside a big room.

This expectation stands up to scrutiny. Analytical solutions describing an infinitely-thin, spherically symmetric shell
surrounding a static BH were constructed (and shown to be stable in some regions of parameter space~\cite{1990CQGra...7..585F,Brady:1991np}). They are described by the geometry
\beq
ds^2&=&-\alpha\left(1-\frac{2M_-}{r}\right)dt^2+\left(1-\frac{2M_-}{r}\right)^{-1}dr^2+r^2d\Omega^2\,,\quad r<R\,,\nonumber\\
ds^2&=&-\left(1-\frac{2M_+}{r}\right)dt^2+\left(1-\frac{2M_+}{r}\right)^{-1}dr^2+r^2d\Omega^2\,,\quad r>R\,,
\eeq
with $\alpha=\frac{1-2M_+/R}{1-2M_-/R}$ and where $R$ stands for the radius of the shell, which is described by an energy surface density $\sigma$ and pressure $P$, related through
\be
k_--k_+=4\pi R\sigma\,,\qquad \frac{M_+}{k_+}-\frac{M_-}{k_-}=4\pi R^2(\sigma+2P)\,,
\ee
with $k_\pm=(1-2M_\pm/R)^{1/2}$. A finite-thickness shell version was studied recently~\cite{Vogt:2009gs,Vogt:2010ad}.

The previous solution is somewhat artificial, in that the shell of matter is infinitely thin, and supported outside the horizon. However, even reasonable ``short-hair'' (``short'' because it can be localized arbitrarily close to the horizon) solutions are possible, and can be found in closed analytical form.
An example of a BH solution surrounded by an anisotropic fluid is described by~\cite{Brown:1997jv},
\beq
ds^2&=&-f dt^2+\frac{dr^2}{f}+r^2d\theta^2+r^2\sin^2\theta\,d\phi^2\,\nonumber\\
f&=&1-\frac{2M}{r}+\frac{Q_m^{2k}}{r^{2k}}\nonumber\\
\rho&=&\frac{Q_m^{2k}(2k-1)}{8\pi r^{2k+2}},\,\quad P=k\rho\nonumber
\eeq
where $\rho$ and $P$ are the density and (angular) pressure, respectively, of the anisotropic fluid, and $Q_m$ is a constant, describing the ``matter-hair''. The stress-energy tensor of the fluid is given by
\be
T_{\mu\nu}=\rho U_{\mu}U_{\nu}+P\sigma_{ab}-\rho u_{\mu}u_{\nu}\,,
\ee
where $U_\mu$ is the fluid four-velocity, $u_\mu$ is a unit radial vector such that $u^\mu U_\mu=0$ and $\sigma_{\mu\nu}=g_{\mu\nu}+U_{\mu}U_{\nu}-u_{\mu}u_{\nu}$.
The solution above corresponds to an anisotropic stress-tensor, specified by (3.2) in Ref.~\cite{Brown:1997jv}, and it reduces to the Reissner-Nordstrom BH when $k=1$.

It seems unlikely that these hairy solutions form in practice, in any kind of realistic environment. However, hairy solutions that {\it have} been observed in numerical simulations are compact torii around the final BH formed as the endstate of the coalescence and merger of two neutron stars~\cite{Rezzolla:2010fd}. These torii can have masses as large as $10\%$ of the total mass, and seem to be stable on dynamical timescales.
\subsection{Hairy black holes in the Standard Model and extensions thereof\label{subsec:sm}}

The above solutions refer to fluids, which are themselves effective descriptions of fermions, and have not
- with the exception of torii - been observed to form in realistic scenarios. We will now discuss a more natural framework for hairy BHs, in the context of {\it fundamental fields}. We will not try to make a systematic classification of solutions in this setup, and instead refer the reader to some of the outstanding reviews out there~\cite{Heusler:1996ft,Chrusciel:2012jk,Berti:2015itd,Herdeiro:2015waa,Sotiriou:2015pka,Volkov:2016ehx}. We would like to stress however, that all solutions need massive or self-interacting fields, in line with the fluid counterpart discussed previously.

Conceptually, the discovery of BHs with ``color'' (static BH solutions in Einstein-Yang Mills (EYM) theory~\cite{Bizon:1990sr,Volkov:1990sva}, that require for their complete specification an additional parameter -- besides the mass -- not associated with any
conserved charge), was the first example showing that the no-hair hypothesis needs revision. The theory is
EYM with SU(2) gauge group, described by the action 
\be
S=\int d^4x\sqrt{-g}\left(R/16\pi G-F^a_{\mu\nu}F^{a\mu\nu}/g^2\right)\,,
\ee
where $F^a_{\mu\nu}$ is the Yang-Mills field strength and $1/g^2$ the coupling constant.
To see how new solutions are possible, let's go back to the perturbative framework of Section~\ref{sec:Sch} and ``freeze'' gravity 
by sending $1/g^2\to 0$. Then, gravity effectively decouples from the YM field, and any vacuum solution solves the Einstein equations.
Let us take again a Schwarzschild background. The YM field can be expressed in terms of a field $\Psi(t,r)=\psi(r)e^{-i\omega t}$. 
The radial wavefunction $\psi(r)$ satisfies the equation \cite{Choptuik:1999gh,Bizon:2007xa}
%
%
%
\be
f\left(f\psi'\right)'+\left(\omega^2+f\frac{1-\psi^2}{r^2}\right)\psi=0\,,\label{eq:YM_Schwarzschild}
\ee
with $f=1-2M/r$. Note that trivial static solutions to this equation are $\psi=0,1$.
Equation (\ref{eq:YM_Schwarzschild}) is stable on the $\psi=1$ branch: its linearized (around $\psi=1$) version is identical to that of $l=1$ electromagnetic modes, and a fundamental frequency $M\omega=0.248263-i0.0924877$ satisfies the necessary boundary conditions~\cite{Bizon:2007xa}. On the other hand, linearization around $\Psi=0$ yields unstable solutions, as is easy to prove. We find the unstable mode $M\omega=i0.1232877$. This indicates that there is a nontrivial static ($\omega=0$) solution of equation (\ref{eq:YM_Schwarzschild}). Such solution can be found imposing the appropriate asymptotic behavior at the horizon and demanding that
$\psi(\infty)=1$.

The violation of the no-hair hypothesis was confirmed with the discovery of static BH solutions 
in theories like Einstein-Skyrme \cite{Bizon:1992gb,Droz:1991cx,Heusler:1991xx}, 
and Einstein-non Abelian-Proca  \cite{Greene:1992fw,Torii:1994nm}. 

As Bekenstein pointed out, these hairy solutions are possible in a way that parallels the existence of charged BHs: the gauge invariance of electrodynamics causes the Coulomb potential to propagate instantaneously in an appropriate gauge. Thus, the argument that information about hair cannot leave the horizon because it would need to travel faster than light, does not apply to the Coulomb potential, and charged BHs exist. Likewise, it seems intuitive that gauge invariance of non-abelian gauge theories should allow one or more of the gauge field components generated by sources in a BH to ``escape'' from it~\cite{Bekenstein:1996pn}.

\begin{figure}
\begin{center}
\begin{tabular}{c}
\includegraphics[width=0.5\textwidth]{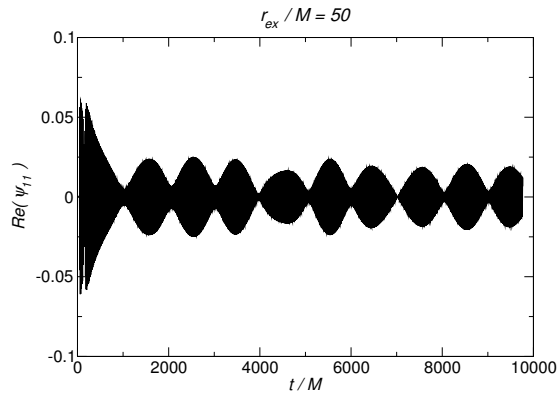} \\
\end{tabular}
\end{center}
\caption{\label{fig:waveform_massive}
Evolution of (the dipole component of) a scalar Gaussian wavepacket in a background Kerr geometry, with $a=0.99M$.
The scalar has mass parameter $M\mu_S = 0.42$. From Ref.~\cite{Witek:2012tr}.
}
\end{figure}
It is possible that a different mechanism might act to allow hairy BHs to exist: instead of allowing the hair to leave the horizon,
one could in principle {\it prevent} it from falling in, in the first place!
Superradiance in BH physics allows for precisely this, amplifying low-frequency bosonic waves, at the expense of the ``horizon's rotational energy''~\cite{Brito:2015oca}.
For massive fields, a finite-height barrier exists at large distances, confining the field and triggering an instability.
Thus Kerr BHs are unstable against massive bosonic perturbations~\cite{Damour:1976kh,Detweiler:1980uk,
Cardoso:2005vk,Dolan:2007mj,Brito:2015oca}. To be specific, the theory of a minimally coupled scalar with a mass term $\mu_S$    
\be
S=\int d^4x \sqrt{-g} \left( \frac{R}{\kappa} 
-\frac{1}{2}g^{\mu\nu}\bar{\Psi}^{}_{,\mu}\Psi^{}_{,\nu} -\frac{\mu_S^2\bar{\Psi}\Psi}{2}\,
\right)\,.\label{minimally_coupled}
\ee
admits Kerr BHs solutions which are linearly unstable against fluctuations of the scalar $\Psi$.
The nonlinear development of the instability is not yet known~\cite{Witek:2012tr,Okawa:2014nda,Zilhao:2015tya},
but linearized evolutions show that sufficiently small dimensionless coupling $M\mu_S$ leads to configurations that can live longer than a Hubble time around a Kerr BH. An example of the evolution of a Gaussian wavepacket around a Kerr BH is shown in Fig.~\ref{fig:waveform_massive}.
Notice how differently massless and massive fields behave (compare with Fig.~\ref{fig:MFScaMLwaveform}).
As such, even minimally coupled massive scalars produce what for all purposes are hairy solutions.

It seems intuitive that for {\it real} fields, the instability acts in such a way as to reduce the angular momentum off the BH, producing a slowly-spinning Kerr BH~\cite{Brito:2014wla}. This is also a consequence of {\it Theorem 3} in Section~\ref{subsec:theorems}.
Complex fields on the other hand, may give rise to a time-independent stress-energy tensor and therefore avoid energy loss, 
and both {\it Theorem 1} and {\it Theorem 2}.
It is thus not surprising that new solutions might branch off Kerr. 
These solutions were described in Refs.\cite{Herdeiro:2014goa,Herdeiro:2015waa} and are the perfect (because of their simplicity) example of how no-hair theorems can be circumvented. 
These ``hairy'' solutions can be generalized to complex vectors~\cite{Herdeiro:2016tmi} (for which the Kerr geometry is also linearly unstable~\cite{Pani:2012bp,Pani:2012vp,Witek:2012tr,Brito:2015oca}) and also to self-interacting scalars~\cite{Kleihaus:2015iea,Herdeiro:2015tia}.

Models of minicharged dark matter predict the existence of new fermions which possess a fractional electric charge or are charged under a hidden $U(1)$ symmetry~\cite{DeRujula:1989fe,Perl:1997nd,Holdom:1985ag,Sigurdson:2004zp,Davidson:2000hf,McDermott:2010pa}. Their corresponding charge is naturally much smaller than the electron charge and their coupling to the Maxwell sector is suppressed. The following classical Lagrangian captures these theories~\cite{Holdom:1985ag} 
\begin{equation}
 {\cal L} = \sqrt{-g}\left(\frac{R}{16\pi}- \frac{1}{4}F_{\mu\nu} F^{\mu\nu}-\frac{1}{4}B_{\mu\nu} B^{\mu\nu}+ 4\pi e j_{\rm em}^\mu A_\mu+ 4\pi  e_h j_h^\mu B_\mu +4\pi  \epsilon e j_h^\mu A_\mu\right)\,,\label{Lagrangian}
\end{equation}
where $F_{\mu\nu}:=\partial_\mu A_\nu-\partial_\nu A_\mu$ and $B_{\mu\nu}:=\partial_\mu B_\nu-\partial_\nu B_\mu$ are the field strengths of the ordinary photon and of the dark photon, respectively, $j_{\rm em}^\mu$ and $j_h^\mu$ are the electromagnetic and the hidden number currents, $e$ is the electron charge, and $e_h$ is the gauge coupling of the hidden sector. The model~(\ref{Lagrangian}) describes a theory in which a charged fermion is coupled to ordinary photons with coupling $\epsilon^2 e^2$ and to dark photons with coupling $e_h^2:=\epsilon_h^2 e^2$. The parameters $\epsilon$ and $\epsilon_h$ are free.
In this theory, BHs are described by the Kerr-Newman family of geometries, but now the classical and quantum discharge mechanisms can be suppressed. Thus, BHs can acquire electric ``hair''~\cite{Cardoso:2016olt}.

Finally, another possible mechanism for hair growth, consists on dropping the stationarity assumption, and use time-dependent boundary conditions. Under certain conditions, this can lead to non-trivial BH geometries~\cite{Jacobson:1999vr,Berti:2013gfa,Babichev:2013cya}.


\subsection{Hairy black holes in other theories or frameworks}

In addition to simple {\it extensions} of GR, there are a number of proposed modifications of GR that actually change
the gravity sector. The ``zoo'' is too big to describe here, we refer the reader to recent reviews on the
topic~\cite{Berti:2015itd,Yagi:2016jml}. Some modifications are string-theory motivated and are also expected to arise
in a completion of GR, and in this sense it is not too surprising to find theories which include higher-order-in-curvature terms, like Chern-Simons gravity~\cite{Alexander:2009tp} or Einstein-dilaton Gauss-Bonnet (EDGB)
gravity~\cite{Mignemi:1992nt,Kanti:1995vq}. These theories give rise to hairy BH
solutions~\cite{Mignemi:1992nt,Kanti:1995vq,Yunes:2009hc,Konno:2009kg,Sotiriou:2013qea}.
We should also point out that even scalar-tensor theories can give rise to BHs surrounded by scalar fields, although the scalar fields needs nontrivial
matter content to be anchored on~\cite{Cardoso:2013fwa,Cardoso:2013opa}.

Other modifications of GR are motivated by a search for massive gravitons, a quest related to solutions of the
cosmological constant problem, but also to nonlinear completions of GR. BHs in some of these theories also
circumvent uniqueness and no-hair results and may be surrounded by massive-graviton
hair~\cite{Brito:2013xaa,Babichev:2015xha,Volkov:2016ehx}. Hairy BH solutions also appear in
Lorentz-violating gravity theories. BH solutions naturally have hair in these theories, because the preferred foliation
can be described in terms of a scalar field (see e.g.~\cite{Barausse:2013nwa} and references therein).


Recent studies of quantum effects in BH geometries argue that BHs should be surrounded
by ``soft'' hair at quantum level~\cite{Hawking:2016msc}, while some advocate that hair is a rule rather than the exception~\cite{Gubser:2005ih}.

To conclude, BHs acquire hair in a variety of setups.
\subsection{Horizonless compact objects}

The no-hair hypothesis in astrophysical settings is tangled with the
assumption that the compact object is too massive and compact to be anything else than a BH.
Theories as simple as gravitating massive scalars or vectors give rise to objects -- {\it boson stars} and {\it oscillatons} -- which could mimic BHs: they can be very massive, they are dark because the interaction cross-section with normal matter is small,
and they can be very compact~\cite{Jetzer:1991jr,Schunck:2003kk,Liebling:2012fv,Macedo:2013jja,Brito:2015pxa,Brito:2015yfh}. It turns out that even the optical appearance of boson stars can be similar to that of BHs~\cite{Vincent:2015xta}.
Other proposals for compact massive objects include gravastars~\cite{Mazur:2004fk}, superspinars~\cite{Gimon:2007ur}, wormholes~\cite{Damour:2007ap} and even mixed wormhole-star systems~\cite{Dzhunushaliev:2016ylj}.
Superspinars invoke the existence of unknown quantum effects that allow the existence of Kerr geometries
without classical horizons.

Very compact objects with a hard and thermally emitting surface, such as gravastars, 
can be strongly constrained by observations~\cite{Broderick:2007ek,Broderick:2005xa}.
However, these constraints assume that the radiation channel is all in some Standard Model particles
and is unlikely to strongly constrain boson stars, for example.

\section{``No-hair'' rises again}

Despite the several conceptual deaths of the no-hair hypothesis, there are good reasons to believe that
astrophysical BHs are -- to a good extent -- described by the Kerr geometry:
  
\begin{itemize}

\item When a BH is surrounded by an accretion disk, the density of the disk is so small that the deviations from Kerr
  spacetime are tiny and, in many respects, can be neglected. Typically the disk can be understood as moving in a Kerr background geometry and {\it used} to infer properties of that background (as we explain in Section~\ref{sec:accretion}). More massive matter configurations may form (for instance, the
  dense disk discussed in~\cite{Rezzolla:2010fd}), but it is difficult to imagine that they persist for a
  significant fraction of the BH life.
  
\item There is no indication that {\it realistic} collapse scenarios lead to the presence
of substantial amounts of any of the hair discussed previously. 
Evolutions of the superradiant instability, for example (see Section~\ref{subsec:sm}) indicate
that minimally coupled scalar clouds make up {\it at most} $\sim 30\%$ of the system ADM mass, and that it is spread over a wide region~\cite{Brito:2014wla} (therefore with negligible backreaction). 
Likewise, with the exception of boson stars and oscillatons~\cite{Brito:2015yfh,Okawa:2013jba}, none of the other ultracompact horizonless objects seem to arise as endpoints of gravitational collapse.
Some of these objects (like wormholes, etc) are ``cut and paste'' or very contrived constructions, casting doubts on their ability to make actual predictions or doubts that their dynamics can ever resemble closely that of the objects they are supposed to mimick~\cite{Yunes:2016jcc}.

Note also that astrophysical BH-candidates come in all scales, from stellar mass to gigantic, supermassive objects.
Boson stars and oscillatons, on the other hand, have a maximum mass (and compactness) dictated by the mass of some fundamental boson. It is thus hard to devise natural mechanisms to explain all observations.

\item Even though hairy BH solutions or horizonless compact objects exist as equilibrium solutions of the field equations,
some (many?) are dynamically unstable in a large portion of the parameter space: BHs in EYM theory are unstable~\cite{Straumann:1990as}; 
BHs surrounded by minimally coupled massive scalars are suspected to be unstable in parts of the parameter space against an ergoregion instability~\cite{Herdeiro:2015gia,Brito:2015oca};
horizonless, spinning ultracompact objects\footnote{by which it is meant an horizonless object with a light ring.} are unstable also against ergoregion instabilities~\cite{Cardoso:2007az,Pani:2009fd,Pani:2010jz}.
In fact, some studies also suggest that {\it any} ultracompact object (spinning or not) is unstable~\cite{Keir:2014oka,Cardoso:2014sna}\footnote{Curiously, all the stable configurations of a BH surrounded by a thin spherical shell of anisotropic fluid, described in \ref{subsec:fluid}, require that $R>3M_-$, i.e., that the shell is outside the BH circular light ring. In this sense, it conforms to a ``no short-hair'' theorem~\cite{Nunez:1996xv}, but can also be looked at as an example that ultracompact objects are generically unstable.}.

\item It is possible that a carefully concocted horizonless object is stable on small timescales. However, for compactnesses extremely close to that of a BH (for example by modifying the geometry on Planck scales),
the geometry is for many purposes still that of a BH in GR. In fact, such objects might be very hard to distinguish, observationally, from BHs~\cite{Damour:2007ap,Barausse:2014tra,Cardoso:2016rao}.

\item In a large class of modified or extended theories of gravity, such as scalar-tensor or $f(R)$ theories, the Kerr geometry is still an equilibrium solution~\cite{Psaltis:2007cw}, albeit not necessarily unique.
Insofar as tests of the background geometry are concerned, they are indistinguishable from 
GR BHs, but their fluctuations do also probe the underlying theory~\cite{Barausse:2008xv}, making GW-based tests of the no-hair hypothesis relevant (see below for an expanded discussion on this point).

\item Modified theories of gravity are typically parametrized by ``small'' coupling parameters, which induce parametrically small changes in the corresponding BH solutions. In addition, and perhaps even more relevant, ultracompact objects or Kerr BHs surrounded by small amounts of matter may be observationally indistinguishable from Kerr BHs: their ringdown waveform depends mostly on the properties of the unstable null circular geodesic, which is typically not significantly affected by ``dirtiness'' nor even by the presence or absence of an horizon~\cite{Barausse:2014tra,Cardoso:2016rao}.

\end{itemize}


All the arguments above indicate that spinning BHs in our universe are likely to be described well by the Kerr
geometry. The no-hair hypothesis is -- and should be -- taken seriously
In any case, observations hinting otherwise
would provide clear signs of new physics or fields, and therefore measurements of ``hair'' are pursued with vigour.

\section{Tests of the ``no-hair'' hypothesis}

In order to test whether a compact object is described by the Kerr geometry, one needs to make observations. These can
be, for instance, observations of stars orbiting around the compact dark companion, or observations of GWs from its oscillations. The former is an example of a {\it non-dynamical} test, which probes whether the stationary
spacetime metric of the compact object is described by the Kerr solution.  The latter, instead, is an example of a {\it
  dynamical} test, since it probes the compact object behaviour in a dynamical process. 

\subsection{Multipole moments}\label{sec:multipoles}

One of the most natural ways to test the spacetime metric of a compact object is to study - through astrophysical or
GW observations - the motion of stellar objects in its surroundings. We here discuss the {\it
  multipole expansions} framework, which is probably the most appropriate to describe and perform these (non-dynamical)
tests of the ``no-hair'' hypothesis. Multipole expansions have been first introduced in Newtonian mechanics, to describe the gravitational (or electrostatic) 
potential generated by a distribution of masses (or charges) in terms of a set of scalar quantities, the multipoles (see
e.g., Ref.~\cite{jackson1999classical} and the first chapter of Ref.~\cite{poisson2014gravity}, and references therein). They have
then been extended to GR~\cite{Geroch:1970cd,Hansen:1974zz,Thorne:1980ru,Fodor:1989aa}.

Multipole expansions are a powerful tool to extract physical content from a (gravitational or electrostatic) potential,
or from a spacetime metric. Indeed, the multipoles capture all the properties of the potential or of the metric, and in
many cases the observable quantities can be expressed in terms of a multipolar decomposition.  The measurement of the
different multipole moments, through electromagnetic and GW observations, would allow the mapping of a BH spacetime.

In the following, we shall briefly introduce multipolar expansions in Newtonian gravity and in GR; further details are
discussed in the Appendix.

\subsubsection{Multipolar expansions in Newtonian gravity and in general relativity}\label{sec:multipolesngr}

\begin{figure}[htbp!]
\centering
\includegraphics[width=0.60\textwidth]{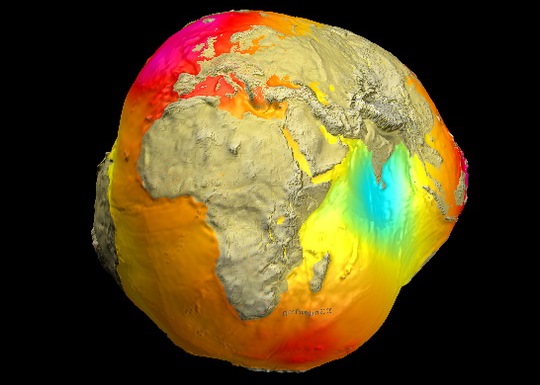}
\caption[]{The gravitational field of the Earth (known as the Potsdam Potato), based on data from the LAGEOS, GRACE, and
  GOCE satellites and surface data. Gravitational field strength is represented by elevation and color.  Credit: CHAMP~\cite{CHAMP},
GRACE~\cite{GRACE}, Research Center for Geophysics (GFZ)~\cite{GFZ}, NASA~\cite{NASA}, DLR~\cite{DLR}.
\label{fig:Potsdam:potato}}
\end{figure}

In Newtonian gravity, the gravitational potential $\Phi(t,\vec x)$ in the exterior of a body with mass density
$\rho(t,\vec x)$ is the solution - in vacuum - of Poisson's equation $\nabla^2\Phi=4\pi G\rho$. This solution can be
written as a series expansion in $1/r$ (where $r=|\vec x|$), called {\it multipolar expansion of the potential}, which,
in the case of a {\it stationary, axisymmetric body}, is
\begin{equation}
\Phi(\vec x)=-G\sum_{l=0}^\infty\frac{M_l}{r^{l+1}}P_l(\cos\theta)\,,\label{multnewtaxis}
\end{equation}
where $P_l(\cos\theta)$ are the usual Legendre polynomials, and
\begin{equation}
M_l=\int\rho({\vec x})r^lP_l(\cos\theta)d^3x\label{Mlintegral}
\end{equation}
are the {\it multipole moments} of the body. The first moments are well known: $M_0=M$ is the body's mass; $M_1$ is its
dipole moment, vanishing in the center-of-mass frame; $M_2=Q$ is the quadrupole moment, $M_3$ is the octupole moment,
and so on. Introducing dimensionless multipole moments $J_l=-M_l/(MR^l)$, where $R$ is a characteristic length of the
body (in the case of the Earth, $R$ is the equatorial radius),
\begin{equation}
\Phi=-G\left[\frac{M}{r}-J_2\frac{MR^2}{r^3}P_2(\cos\theta)+\dots\right]\,.
\end{equation}
The multipole moments of an astronomical body are a measure of its departure from spherical symmetry. Generally, planets
and stars rotate so slowly that their deviation from spherical symmetry is tiny (other sources of asymmetry are even
smaller). For instance, for the Earth $J_2=1.083\times10^{-3}$, and the $l>2$ dimensionless multipoles are at least a 
thousand times smaller. These quantities have been determined by studying the motion of satellites orbiting around the
Earth~\cite{tapley2004gravity,drinkwater2003goce,ciufolini2012overview}.
A depiction of the gravitational field of the Earth as reconstructed with data from these mission is shown in Fig.~\ref{fig:Potsdam:potato}, where differences
in the gravitational field across the globe are 
represented by elevation and color.

The generalization of multipolar expansions to GR is not straightforward, due to the non-linearity of Einstein's
equations. The {\it relativistic multipole moments} of a {\it stationary, asymptotically flat} spacetime have been
defined by Geroch and Hansen~\cite{Geroch:1970cd,Hansen:1974zz}, with a complex mathematical construction that allows to
describe the deviation of the asymptotic geometry from flatness in terms of two sets of tensorial quantities evaluated
at the point at infinity: the {\it mass multipole moments} and the {\it current multipole moments}
(see the Appendix). In the Newtonian limit, the mass multipole moments reduce to the moments in Newtonian
theory~\footnote[1]{The current multipole moments do not appear in Newtonian theory, because they do not affect the motion
  of masses.}. It has been shown that the spacetime is uniquely determined by its multipole
moments~\cite{beig1980proof,beig1981multipole,kundu1981analyticity}: in other words, they completely characterize the
spacetime geometry outside any stationary body. It has also been shown (in the axisymmetric case) that it is possible to
reconstruct the full spacetime from any ``well-behaved'' set of relativistic multipole
moments~\cite{Backdahl:2005uz,Backdahl:2006ed}.

As in the Newtonian case, when the body (and the spacetime) is axisymmetric, the relativistic (mass and current)
multipoles reduce to a set of scalar quantities $(M_l,J_l)$~\cite{Hansen:1974zz,Fodor:1989aa}; if the spacetime is
reflection-symmetric (i.e., symmetric with respect to a reflection on the equatorial plane), the odd mass moments and
the even current moments identically vanish: $M_{2l+1}=S_{2l}=0$. As in the Newtonian case, $M_0=M$ is the mass, $M_2=Q$
is the quadrupole moment; moreover, $S_1=J$ is the angular momentum. For the Kerr spacetime,
\begin{eqnarray}
M_{2l}&=&(-1)^{l}Ma^{2l}\nonumber\\
S_{2l+1}&=&(-1)^{l}Ma^{2l+1}\,.\label{Kerrmoments}
\end{eqnarray}
Notice that this expression is a manifestation of (``no-hair'') {\it Theorem 1} (Sec.~\ref{subsec:theorems}), as it fixes all multipole moments as function of two parameters only.

An alternative definition of relativistic multipole moments of stationary, asymptotically flat spacetimes has been
introduced by Thorne~\cite{Thorne:1980ru}. This is an extension of the standard procedure of extracting the mass and the
angular momentum from the far-field limit of the spacetime metric~\cite{MTW}. In Thorne's construction, all multipole
moments can be read out from the asymptotic spacetime metric. In the axisymmetric case
\begin{eqnarray}
g_{00}&=&-1+\frac{2M}{r}+\sum_{l\ge2}\frac{2}{r^{l+1}}\left[M_lP_l(\cos\theta)+(l'<l~\hbox{harmonics})\right]\nonumber\\
&=&-1+\frac{2M}{r}+\frac{2Q}{r^3}P_2(\cos\theta)+\dots\nonumber\\
g_{0\phi}&=&-2\sin^2\theta\sum_{l\ge1}\frac{1}{r^l}\frac{S_l}{l}\left[P'_l(\cos\theta)+(l'<l~\hbox{harmonics})
\right] \nonumber\\
&=&-\sin^2\theta\left(\frac{2J}{r}+\frac{2S_3}{3r^3}P'_3(\cos\theta)+\dots\right)\,.\label{tmoments}
\end{eqnarray}
This definition requires that the coordinate system belongs to a special class, the ``asymptotically Cartesian and mass
centered'' (ACMC) coordinates, which means that it becomes Minkowski with sufficient rapidity at large radii
(see the Appendix for details), and that the origin of the space coordinates lies at the center of mass of the
source. As long as the coordinates are ACMC, the multipole moments $(M_l,S_l)$ are coordinate-independent. Moreover, it
has been shown that the definitions of relativistic multipole moments given by Geroch-Hansen and by Thorne are actually
coincident~\cite{gursel1983multipole}~\footnote[2]{Note that Thorne~\cite{Thorne:1980ru} uses a normalization for
  multipole moments different than that of Geroch and Hansen~\cite{Geroch:1970cd,Hansen:1974zz}. In
  Eqns.~(\ref{tmoments}) we follow the conventions of Geroch-Hansen, which is also adopted in most of the recent
  literature on the subject.}.

In the case of a weak-field source, the multipole moments of a stationary spacetime can also be expressed as integrals
over the source of the energy and momentum densities~\cite{mathews1962gravitational}, i.e. (in the axisymmetric case) by
Eq.~(\ref{Mlintegral}) for mass multipoles, and a similar expression for current
multipoles~\cite{Ryan:1996nk,Stein:2013ofa}. This definition can be extended, using properly defined ``effective''
(energy and momentum) densities, to the case of strong-field sources, as long as the source can be covered by a
so-called ``de Donder'' coordinate frame~\cite{Thorne:1980ru}; this is the case, for instance, of compact
stars. Multipole moments of BHs, instead, can only be defined in terms of their asymptotic geometry.

Given a spacetime metric expressed in a specific coordinate frame, there is a simple method to compute the Geroch-Hansen
multipole moments of the spacetime. This procedure is based on Ryan's formula~\cite{Ryan:1995wh}, which was introduced
as a phenomenological tool to express observable quantites in terms of multipole moments (and will be discussed in more
detail in Sec.~\ref{sec:Ryan}), but is also a powerful {\it computational tool}. One computes the energy of (geodesic)
circular orbits as a function of the orbital frequency $E(\nu)$. A particle moving in that geodesic would emit GWs at
frequency $f=2\nu$, and the so-called ``gravitational spectrum'', i.e. the amount of GW energy emitted per logarithmic
interval of frequency, is $\Delta E(f)=fdE_{gw}/df=-\nu dE/d\nu$.  Ryan has found a general, explicit expression of
$\Delta E(f)$ in terms of the multipole moments of the spacetime (Eq.~(\ref{deltaEmom}) in Sec.~\ref{sec:Ryan});
comparing the function $\Delta E(f)$ with this expression, one can extract the entire set of (Geroch-Hansen) multipole
moments. This approach can be useful to express the asymptotic expansion of the metric - which in general in not
expressed in an ACMC coordinate frame - in terms of the gauge-invariant multipole moments, for instance when the metric
is the result of a numerical integration~\cite{Laarakkers:1997hb,Berti:2003nb,Pappas:2012ns,Yagi:2014bxa}, or of a
perturbative computation~\cite{Pani:2015hfa,Pani:2015nua}. When the metric is known in ACMC coordinates, instead, the
multipoles can be extracted by comparison with Thorne's expansion~(\ref{tmoments}), as e.g. in~\cite{Ayzenberg:2014aka}.

\subsubsection{The multipole moments of some astrophysical objects}
%
\begin{figure}[htbp!]
\centering
\includegraphics[width=0.7\textwidth]{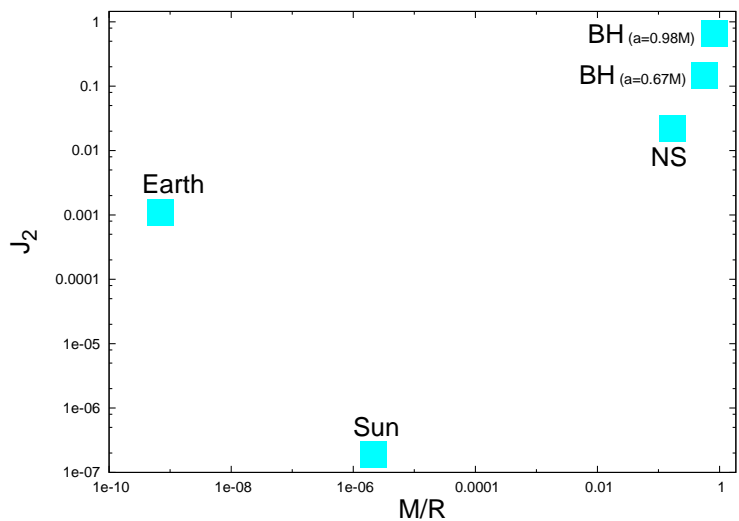}
\caption{Typical values of the normalized quadrupole moments for the Earth, the Sun, a neutron star and two examples of BH.  The
  neutron star~\cite{Benhar:2005gi,Pappas:2012qg} has a mass $M=1.4\,M_\odot$, a radius $R=12$ km, is spinning at $716$
  Hz (the maximum observed spin rate~\cite{hessels2006radio}) and is described using the equation of state of
  Ref.~\cite{Akmal:1998cf}. The BHs are the final BH in the GW150914 coalescence~\cite{Abbott:2016blz}, having $a=0.67M$, and a
  near-extremal BH with $a=0.98M$.
\label{fig:multipoles}}
\end{figure}
To conclude this discussion, we show in Figure~\ref{fig:multipoles} typical values of the normalized quadrupole moment
$J_2=-Q/(MR^2)$, as a function of the compactness $M/R$, for different astrophysical objects: the Earth and the
Sun~\cite{pitjeva2005high}; a neutron star; 
the final BH observed in the first LIGO detection~\cite{Abbott:2016blz}, having $a=0.67M$ and $R=1.74M$; a near-extremal
BH with $a=0.98M$ (and thus $R=1.20M$).  This figure shows that the normalized quadrupole moment of astrophysical
objects comes in a variety of ranges, and that therefore the BH multipole moments are not accidentally shared by many
other objects.

However, we should also add that (exotic) ultra-compact objects can have multipole moments arbitrarily close to that of BHs.
For example, in the ultra-compact limit, the mass quadrupole moment of gravastars is~\cite{Pani:2015tga}
\be
Q=\frac{J^2}{M}\left(1+\frac{8}{45\log\left(1-2M/R\right)}\right)\,,
\ee
showing that it can be arbitrarily close to that of a BH with the same mass, i.e., $Q=J^2/M$. This somewhat contrived
example shows that the law~(\ref{Kerrmoments}) is not, by itself, evidence for the existence of a horizon, but the
alternatives usually invoke somewhat more exotic physics.

\subsection{Dynamical tests with gravitational waves: ringdown}

\subsubsection{Using two or more modes to test the Kerr hypothesis}
%
\begin{figure}
\begin{center}
\begin{tabular}{c}
\includegraphics[width=0.6\textwidth]{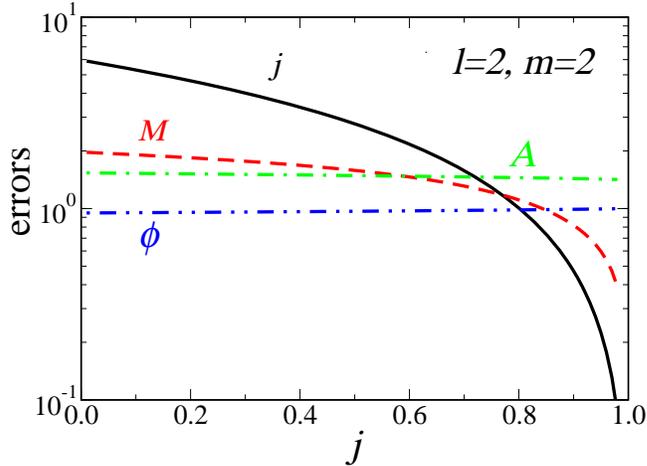} \\
\end{tabular}
\end{center}
\caption{\label{errs-intro}
Errors (multiplied by the signal-to-noise ratio $\rho$)
in measurements of different parameters for the
fundamental $l=m=2$ mode as functions of the
angular momentum parameter $j$. Solid (black) lines give $\rho
\sigma_j$, dashed (red) lines $\rho \sigma_M/M$, dot-dashed (green)
lines $\rho \sigma_A/A$, dot-dot-dashed (blue) lines $\rho
\sigma_{\phi}$, where $\sigma_k$ denotes the estimated rms error for variable
$k$, $M$ denotes the mass of the BH, and $A$ and $\phi$ denote the amplitude and phase of the wave. From Ref.~\cite{Berti:2005ys}.
}
\end{figure}

Tests of the no-hair hypothesis based on the motion and observation of stars at large distances 
probe the {\it background} geometry in the weak-field regime. A multipolar decomposition is meaningful
here, because the contribution of higher multipoles to the motion of stars is suppressed.
Nevertheless, such tests are unable to probe different theories with the same background solution.
To probe the dynamical content of the field equations, dynamical tests are necessary, and GWs
are the ideal tool for this.

As we discussed in Section~\ref{sec:hair_loss}, the vibration modes of BHs in GR are completely determined by the two parameters specifying the Kerr solutions. In any other theory with the same Kerr background, the functional relation among modes would be different.
Thus, {\it conceptually} one can proceed as follows: measure the dominant mode of oscillation, which is characterized by a ringing frequency $\omega$ and a damping time $\tau$. In GR, the dominant mode is invariably (when the oscillations are the result of a BH binary coalescence, which is likely to be the most efficient way to excite the QNMs) that with $l=m=2$, $n=0$ mode, of frequency $\omega_{220}$ and damping time $\tau_{220}$. In the Schwarzschild limit these are the numbers shown in equation~(\ref{eq:QNMs_a0}).
These numbers allow one to determine the (redshifted) mass and angular momentum of the BH. Now measure the most important subdominant mode (typically $l=m=3, n=0$) and check that it is located at the GR prediction. One thus has at hand a null test of GR based on the observation of ringdown modes of BHs.

In practice, there are measurement errors associated with the different sources of noise in the detector, and each of the stages
in the test carries associated uncertainties. These uncertainties are best described in units of the ``signal-to-noise'' ratio (SNR) $\rho$~\cite{Weinstein:1962ig,Scharf:1991ig}. As a rule-of-thumb, a total $\rho\gtrsim 8$ is required for a detection. In the following we focus on the SNR for the ringdown part of the signal only. Fortunately, accurate measurements of the mass and angular momentum of BHs are feasible~\cite{Flanagan:1997sx,Berti:2005ys}: for detection of the fundamental $l=m=2$ bar mode, for example, Figure \ref{errs-intro} shows the estimated
error (multiplied by $\rho$) in measuring the mass $M$, angular momentum parameter $j\equiv a/M$,
QNM amplitude $A_{lmn}$, and phase $\phi_{lmn}$ (see equation (\ref{qnm_exp}) for definitions of these quantities and Ref.~\cite{Berti:2005ys} for further details). At large $\rho$, the uncertainties are well approximated by~\cite{Berti:2005ys,Berti:2007zu}
\be
\sigma_j =\frac{2}{\rho}\left|\frac{Q_{lmn}}{Q_{lmn}'}\right|\,,\qquad \sigma _M =\frac{2}{\rho}\left|\frac{MQ_{lmn}f_{lmn}'}{f_{lmn}Q_{lmn}'}\right|\,, \label{errmEF}
\ee
where primes stand for derivatives with respect to the dimensionless angular momentum $j$, and can be performed with the help of the fits in Eqs.(\ref{ffit})-(\ref{Qfit}) , or taken from the numerical data~\cite{rdweb,Berti:2005ys,Berti:2009kk}.
Even for marginal detection events, the accuracy of mass and spin measurements is at the level of $10\%$
or better.

For large signal-to-noise ratios, one enters the regime where a second mode can be disentangled in the signal.
To understand what criteria needs to be met, consider first disentangling frequencies.
The error associated with the direct measurement of the frequency and damping time of one mode (``1'')
in the signal is ~\cite{Berti:2005ys,Berti:2007zu}
\beq
\rho \sigma_{f_1}&=&\frac{1}{2\sqrt{2}}\left\{\frac{f_1^3\left(3+16Q_1^4\right)}{{\cal A}_1^2 Q_1^7}
\left[\frac{{\cal A}_1^2 Q_1^3}{f_1\left(1+4Q_1^2\right)}+
\frac{{\cal A}_2^2 Q_2^3}{f_2\left(1+4Q_2^2\right)}\right] \right\}^{1/2}\,,\label{sigmaffh}\\
\rho \sigma_{\tau_1}&=& \frac{2}{\pi} \left\{ \frac{\left(3+4Q_1^2\right)}{{\cal A}_1^2 f_1 Q_1} \left[ \frac{{\cal
A}_1^2 Q_1^3}{f_1\left(1+4Q_1^2\right)}+ \frac{{\cal A}_2^2 Q_2^3}{f_2\left(1+4Q_2^2\right)} \right]\right\}^{1/2}\,. \label{sigmataufh}
\eeq
These errors refer to mode ``1'' in a pair. By considering the ``symmetric''
case $\phi_1=\phi_2=0$, the errors on $f_2$ and $\tau_2$ are simply obtained
by exchanging indices ($1\leftrightarrow 2$). 

A natural criterion ({\it \'a la} Rayleigh) to resolve frequencies and damping
times is
\be
\label{criterion} |f_1-f_2|>{\rm max}(\sigma_{f_1},\sigma_{f_2})\,,\qquad |\tau_1-\tau_2|>{\rm
max}(\sigma_{\tau_1},\sigma_{\tau_2})\,. 
\ee
In interferometry this would mean that two objects are (barely) resolvable if
``the maximum of the diffraction pattern of object 1 is located at the minimum
of the diffraction pattern of object 2''. We can introduce two ``critical''
SNRs required to resolve frequencies and damping times,
\be
\rho_{\rm crit}^f =
\frac{{\rm max}(\rho \sigma_{f_1},\rho \sigma_{f_2})}{|f_1-f_2|}\,,\qquad
\rho_{\rm crit}^\tau = \frac{{\rm max}(\rho \sigma_{\tau_1},\rho
\sigma_{\tau_2})}{|\tau_1-\tau_2|}\,,
\ee
and recast our resolvability conditions as
\be
\rho>\rho_{\rm crit}={\rm min}(\rho_{\rm crit}^f,\rho_{\rm crit}^\tau)\,,\qquad
\rho>\rho_{\rm both}={\rm max}(\rho_{\rm crit}^f,\rho_{\rm crit}^\tau)\,. 
\ee
The first condition implies resolvability of either the frequency or the
damping time, the second implies resolvability of both.

\begin{figure}
\begin{center}
\begin{tabular}{c}
\includegraphics[width=0.6\textwidth]{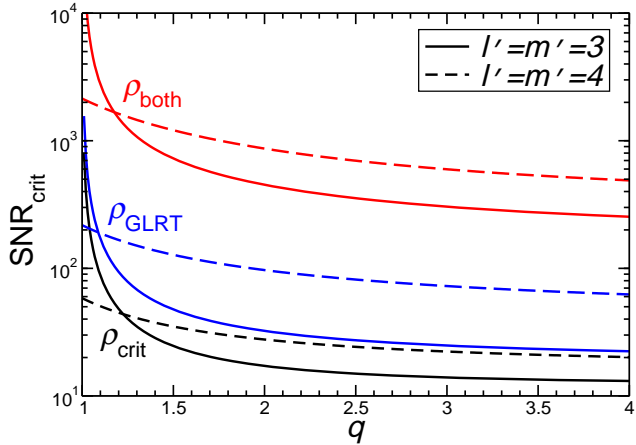} \\
\end{tabular}
\end{center}
\caption{\label{fig:minimumSNR}
Minimum SNR ({\it in ringdown only}) required to resolve two modes, as function of the
  binary's mass ratio $q$. If $\rho>\rho_{\rm GLRT}$ we can tell the presence
  of a second mode in the waveform, if $\rho>\rho_{\rm crit}$ we can resolve
  either the frequency or the damping time, and if $\rho>\rho_{\rm both}$ we
  can resolve both. Mode ``1'' is assumed to be the fundamental mode with
  $l=m=2$; mode ``2'' is either the fundamental mode with $l=m=3$ (solid
  lines) or the fundamental mode with $l=m=4$ (dashed lines). From Ref.~\cite{Berti:2007zu}.
}
\end{figure}
A related question is the magnitude of the signal-to-noise ratio
in order to detect a multi-mode signal, and to resolve two signals of different
{\it amplitudes}. This problem was studied in Ref.~\cite{Berti:2007zu}, and quantified
by deriving a critical signal-to-noise ratio for amplitude resolvability $\rho_{\rm GLRT}$ based on the
{\it generalized likelihood ratio test}. This criteria is almost (but not quite) equivalent to requiring
that the second, sub-dominant mode alone has enough energy that it could be detected on its own.

The different signal-to-noise ratios required to resolve the two modes of a final BH which is the end-product
of an unequal-mass merger of two BHs, is shown in Fig.~\ref{fig:minimumSNR}.
The figure shows that tests of the no-hair hypothesis require large, but not-too-large-to-be-impossible 
signal-to-noise ratios.

\begin{figure*}[h]
\centering
$\begin{array}{cc}
\includegraphics[width=0.45\textwidth]{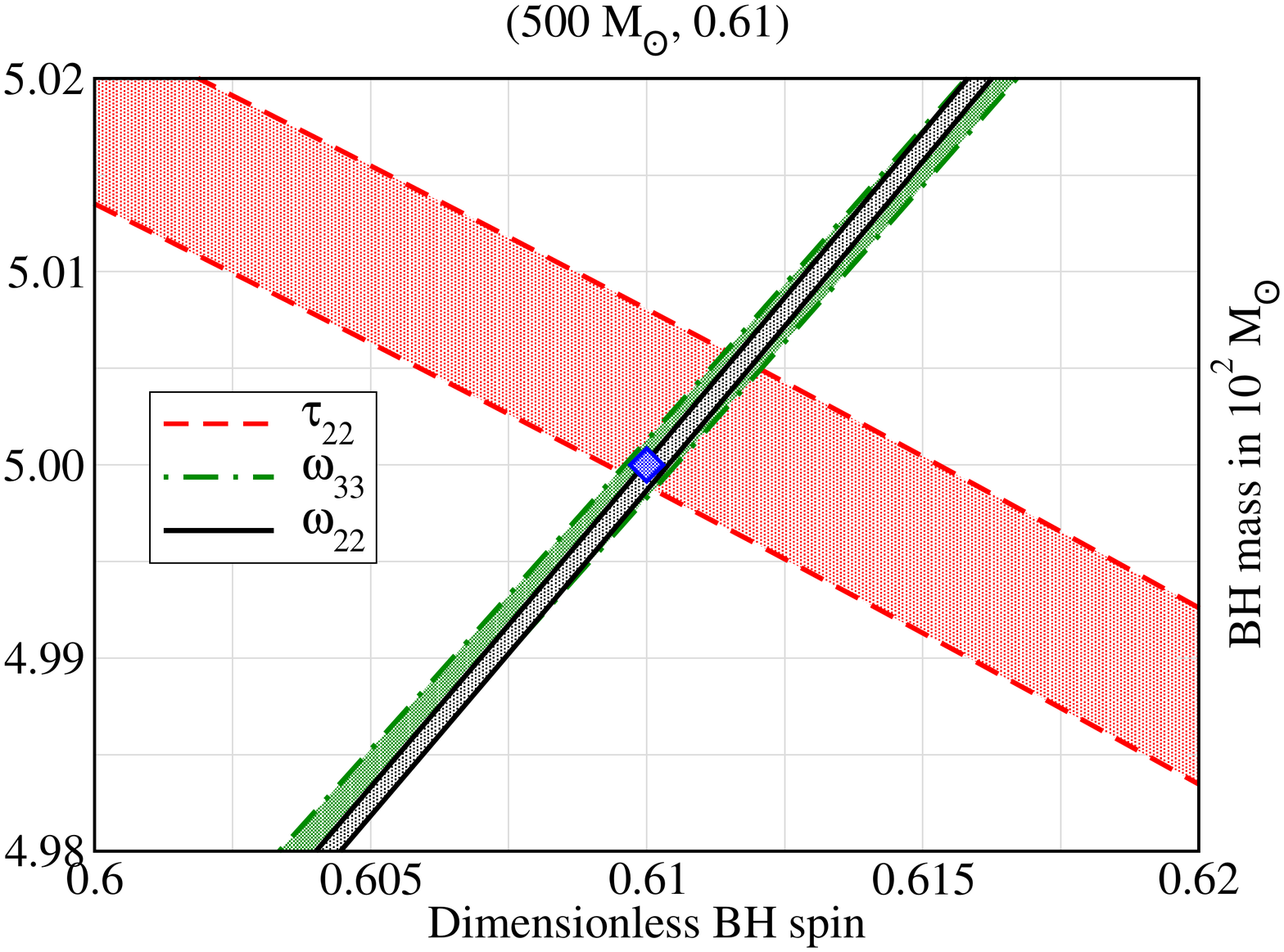} & \includegraphics[width=0.45\textwidth]{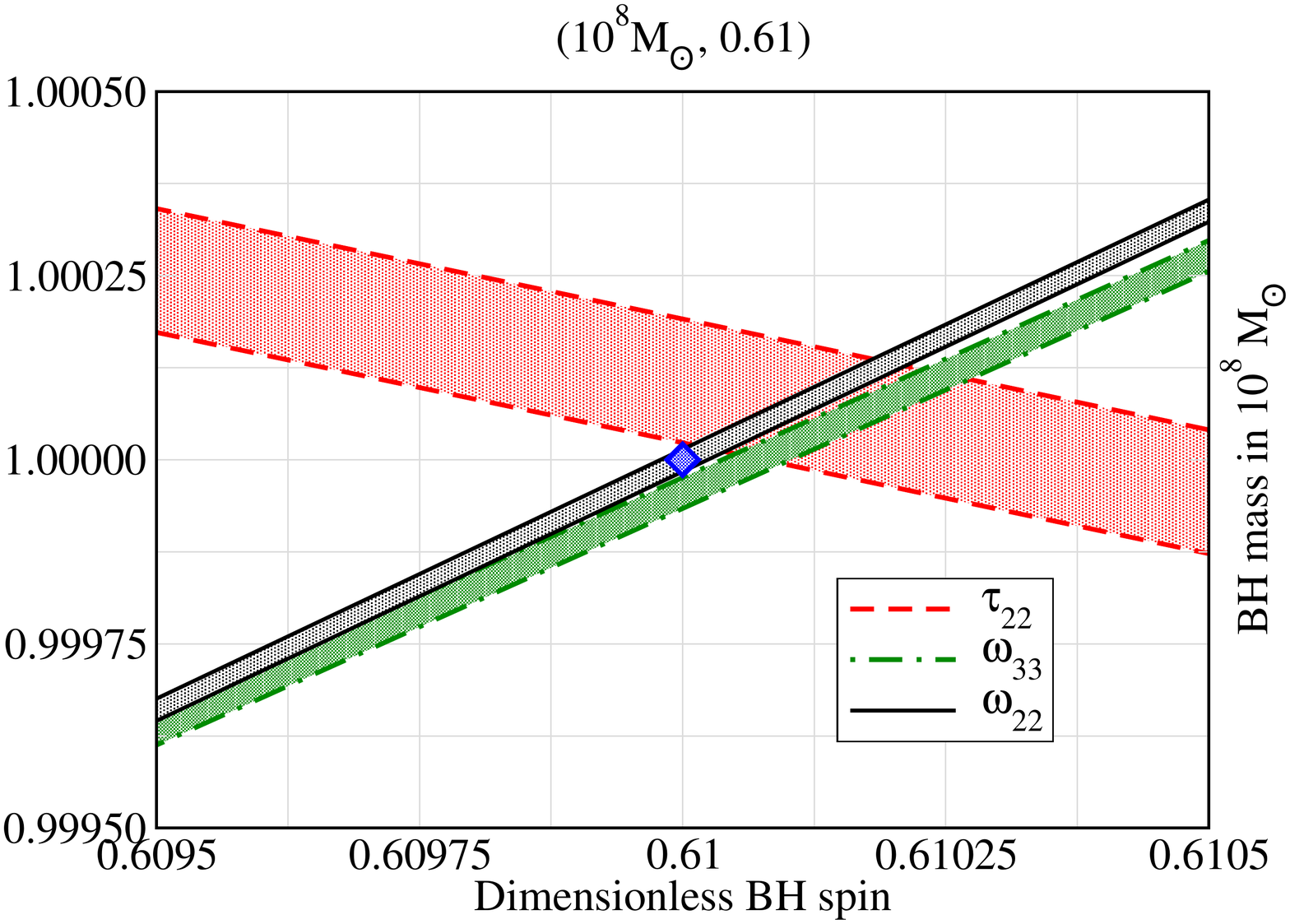}
\end{array}$
\caption{Projections in the $(M, j)$-plane of the 90$\%$ confidence limits on $\omega_{22}$, $\tau_{22}$ and $\omega_{33}$ (blue, blue dotted and red lines respectively)
for injections of signals consistent with GR for $M = 500\,M_{\odot}$ (left at 125\,Mpc; SNR = $2\,888$), 
and $M=10^{8}\,M_{\odot}$ (right at 1\,Gpc; SNR = $115\,154$).  The injected value is denoted in each case by a diamond.
Taken from Ref.~\cite{Gossan:2011ha}.}
\label{minset_GRconfidence}
\end{figure*}
The implementation of this method in an actual search pipeline for two planned GW detectors ET and NGO was described in Ref.~\cite{Gossan:2011ha}. The authors consider a theory-independent parametrization of the QNM frequencies given by
\beq
\omega_{lm}&=&\omega_{lm}^{\rm (GR)}\left(1+\Delta\hat{\omega}_{lm}\right)\\
\tau_{lm}&=&\tau_{lm}^{\rm (GR)}\left(1+\Delta\hat{\tau}_{lm}\right)\,,
\eeq
where $\omega_{lm}^{\rm (GR)}$ are the values of the fundamental QNM frequencies and damping times of Kerr BHs, and 
$\Delta\hat{\omega}_{lm}, \Delta\hat{\tau}_{lm}$ parametrize the relative deviations to these numbers.
The BH is assumed to be the endpoint of a mass ratio $q=2,10$ binary BH merger and its spin is assumed to be $j=0.6,0.26$.

When the signal is described by GR, estimates of mass and spin from, say, three modes will be consistent and yield (generically accurate) measurements of these quantities. This is depicted in Fig.~\ref{minset_GRconfidence}: the three estimates intersect at a common point, which will be the best estimate for the mass and spin of the final BH (the injected signal corresponds to a BH of mass $(500,10^8)\,M_{\odot}$ for NGO and ET respectively, and the best estimate is very close to the injected value).

\begin{figure*}[h]
\centering
$\begin{array}{cc}
\includegraphics[width=0.45\textwidth]{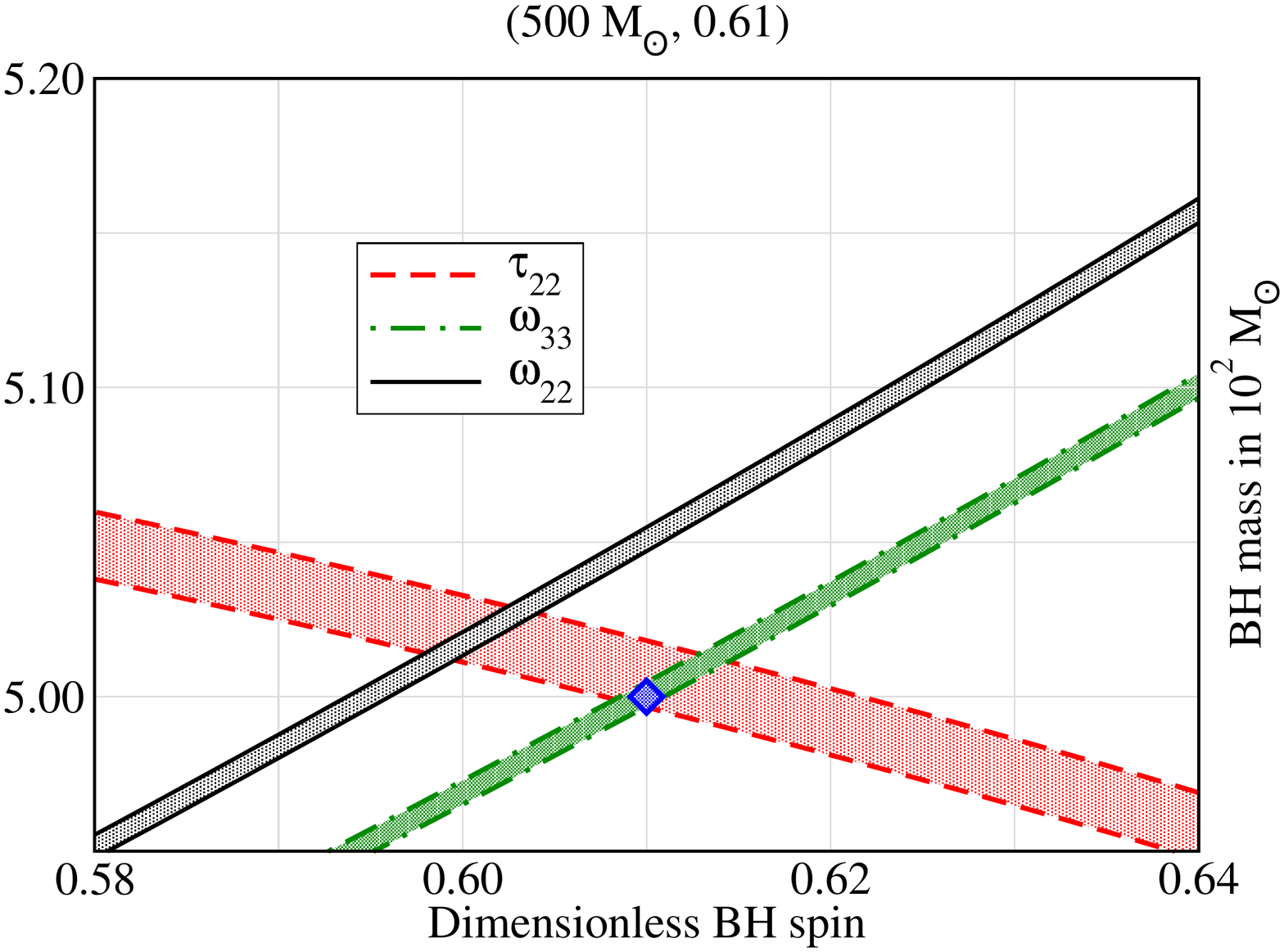} & 
\includegraphics[width=0.45\textwidth]{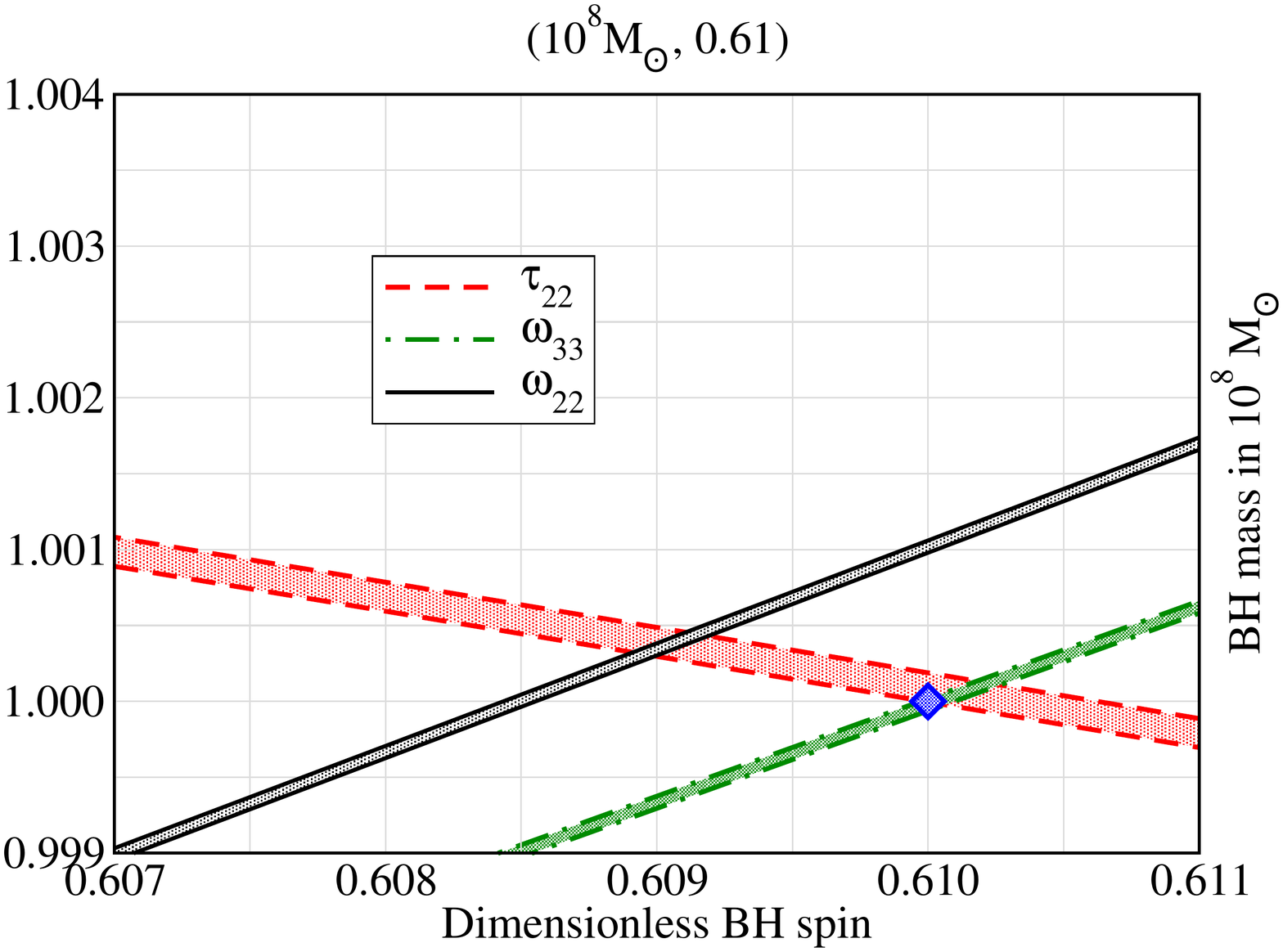}
\end{array}$
\caption{Projections in the $(M, j)$-plane of the 90\% confidence limits on $\omega_{22}$, $\tau_{22}$ and $\omega_{33}$ (blue, blue dotted and red lines, respectively)
for non-GR injections of $M = 500\,M_{\odot}$ (left at 125\,Mpc for the ET telescope; with $\Delta\hat{\omega}_{22} = -0.01$, SNR = $2\,867$) and 
and $M = 10^{8}\,M_{\odot}$ (right at 1\,Gpc for NGO telescope; with $\Delta\hat{\omega}_{22} = -0.001$, SNR = $115\,130$). 
The injected value is denoted in each case by a diamond. Taken from Ref.~\cite{Gossan:2011ha}.}
\label{minset_nonGRconsistency}
\end{figure*}
By contrast, if the signal is inconsistent with GR the intersections of the confidence regions will not agree, as in Fig.~\ref{minset_nonGRconsistency}.

\begin{figure*}[t]
\centering
$\begin{array}{cc}
\includegraphics[width=0.4\textwidth]{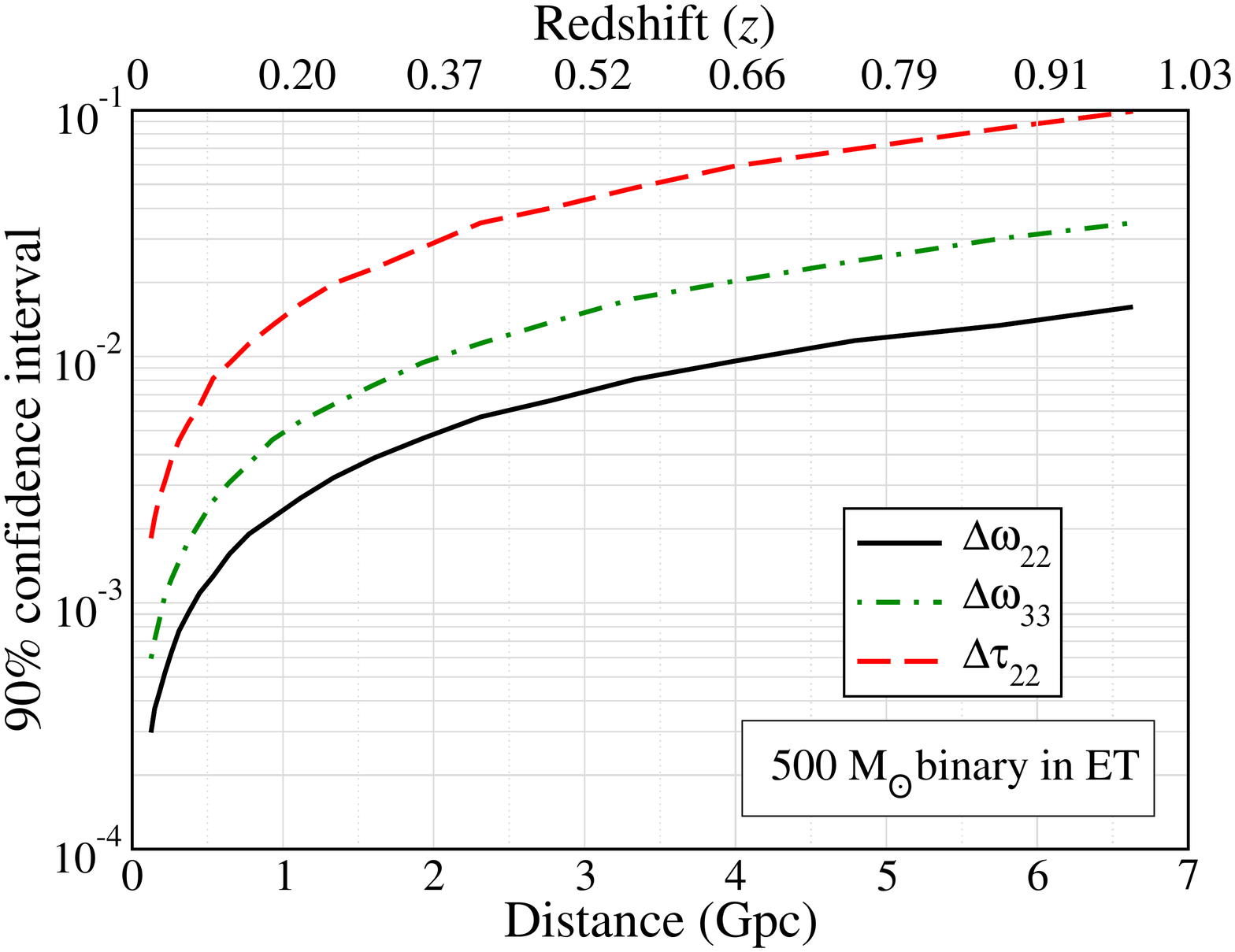}  &
\includegraphics[width=0.4\textwidth]{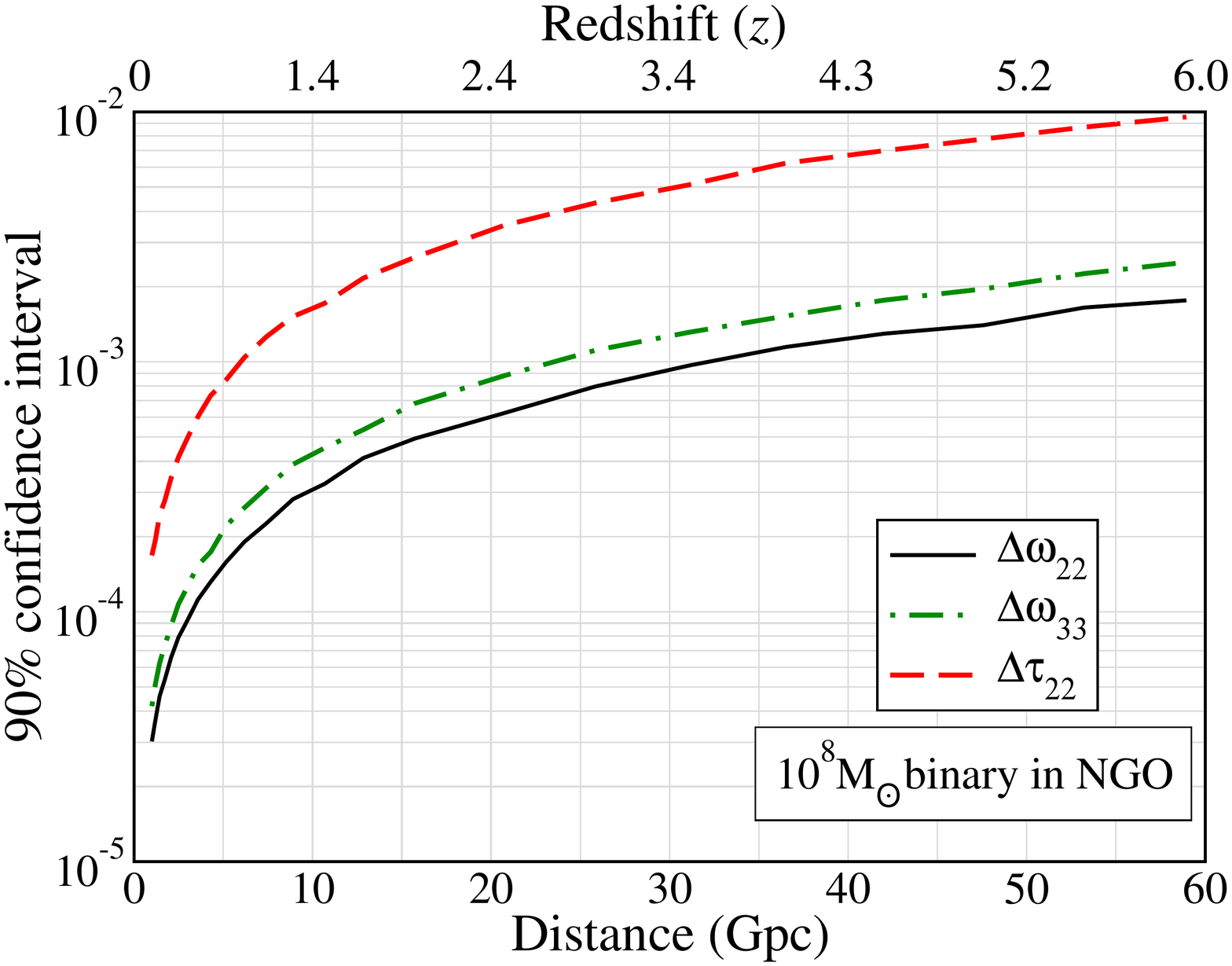} \\
\end{array}$
\caption{Width of the 90$\%$ confidence intervals for $\Delta\hat{\omega}_{22}$,$\Delta\hat{\omega}_{33}$ and $\Delta\hat{\tau}_{22}$ 
(blue, red and blue dotted lines respectively)
against luminosity distance for injections of $500$ (left, ET)
and $10^{8}M_{\odot}$ (right, NGO). Taken from Ref.~\cite{Gossan:2011ha}.}
\label{minset_GRwidths}
\end{figure*}

The accuracy with which the deviation constants can be estimated and the mode parameters resolvable is shown in
Figure \ref{minset_GRwidths}. For each BH system considered, the width of the 90\% confidence intervals for the extracted values of $\Delta\hat{\omega}_{lm}$ and $\Delta\hat{\tau}_{lm}$ were plotted against luminosity distance, for injections with GR waveforms.

\begin{figure*}[!htp]
\centering
  \begin{tabular}{ccc}
    \includegraphics[width=.33\textwidth]{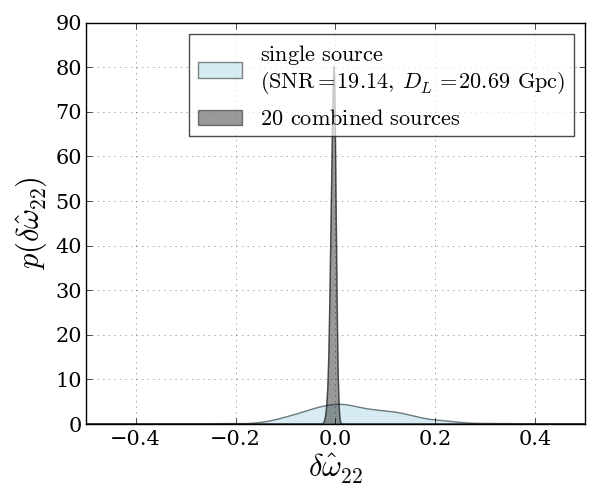} &
    \includegraphics[width=.33\textwidth]{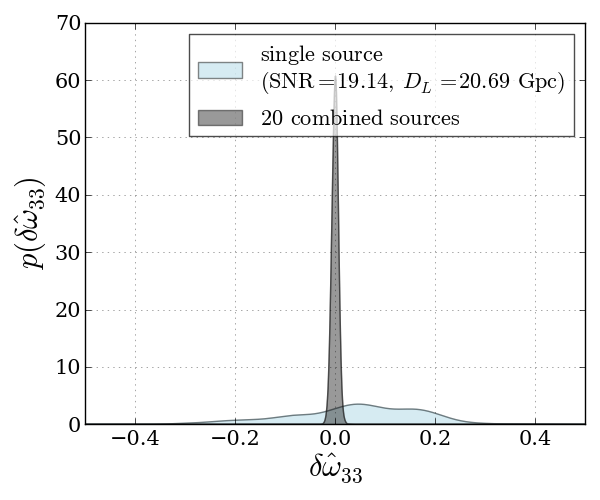} &
    \includegraphics[width=.33\textwidth]{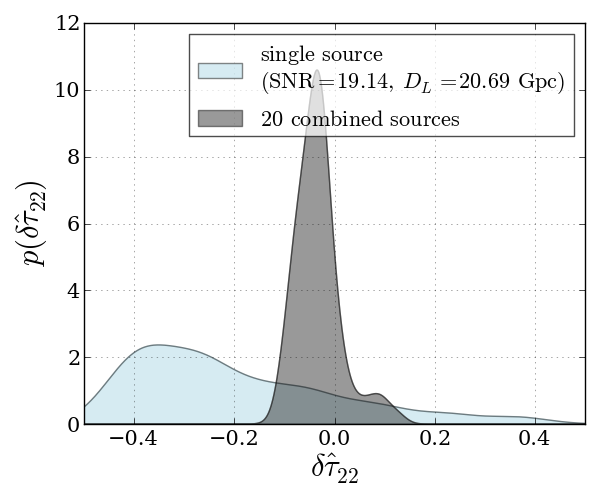} \\
    \includegraphics[width=.33\textwidth]{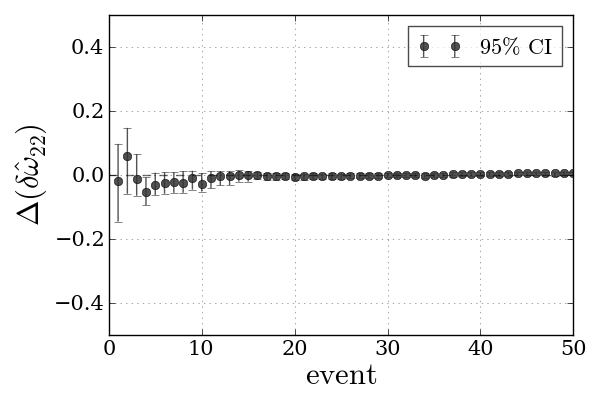} &
    \includegraphics[width=.33\textwidth]{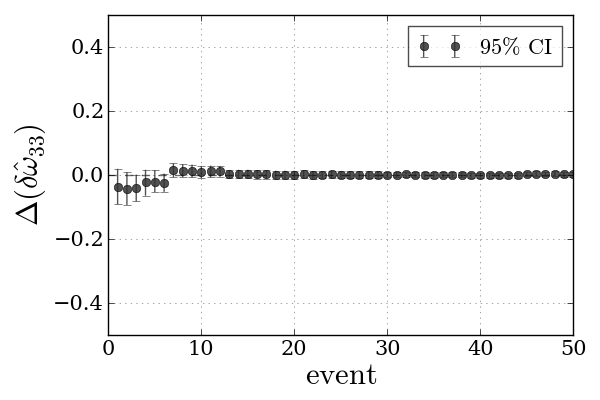} &
    \includegraphics[width=.33\textwidth]{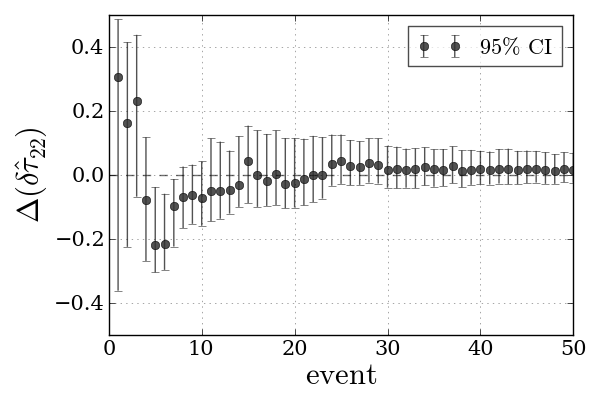}
  \end{tabular}
  \caption{Top panels: Posterior density functions for $\delta\hat{\omega}_{22}$ (left), $\delta\hat{\omega}_{33}$ (middle), and $\delta\hat{\tau}_{22}$ (right), both for a single source at a distance of 20.69 Gpc ($z = 2.47$) with an SNR of 19.14, and for a catalog of 20 sources. Bottom: Evolution of medians and 95\% confidence intervals of PDFs as more and more sources are included.
Taken from Ref.~\cite{Meidam:2014jpa}.}
\label{fig:PDFs}
\end{figure*}
The previous results all assume comparatively large signal-to-noise ratios. 
This analysis was recently extended to multiple detections with larger noise, using a model selection
scheme called TIGER (Test Infrastructure for GEneral Relativity), for the ET telescope~\cite{Meidam:2014jpa}.

As an example, assume that GR is correct ($\delta\hat{\omega}_{22}=\delta\hat{\omega}_{33}=\delta\hat{\tau}_{22}=0$).
How would GW observations of ringdown actually constrain the deviation parameters?
Results are summarized in Fig.~\ref{fig:PDFs}. The top panels show the probability distribution function (PDF) both for an example single source at $D_{\rm L} = 20.69$ Gpc ($z = 2.47$), and for a catalog of 20 sources. As expected, the single-source PDFs are quite wide and uninformative. For $\delta\hat{\omega}_{22}$ and $\delta\hat{\omega}_{33}$, with 20 sources the PDFs become strongly peaked, with very little bias. There is a clear advantage in using all available detections.

If the BH progenitor is known with some accuracy, further tests are possible, even with a single ringdown mode:
the inspiral phase allows, in principle, the mass and spin of the two components to be determined.
GR then predicts a well-defined final BH with a certain mass and spin, a prediction
which can be tested using the dominant ringdown mode~\cite{Ghosh:2016qgn}. In fact, in some cases it might even be possible to {\it infer} the properties of the progenitor from ringdown observations~~\cite{Kamaretsos:2012bs,Kamaretsos:2011um}.

These results were recently used, in conjunction with population synthesis models of the formation and evolution of BH binaries, to estimate how many detections would
yield tests of GR. The results are optimistic for space-based detectors or future generations of Earth-based ones~\cite{Berti:2016lat}.

\subsubsection{Constraining alternative theories}
Up to now, we studied how two modes can test GR. We can turn this around and use measurements of two (or more) ringdown modes to constraint specific modified theories, converting the errors (\ref{sigmaffh})-(\ref{sigmataufh}) on the frequency and damping time to errors on physical quantities by using a simple propagation of errors~\cite{Cardoso:2016olt}. Specifically, 
let $M, j=a/M$ be the mass and dimensionless angular momentum of the BH, and $Q$ an extra parameter which presumably enters the description of the geometry. Since $Q$ measures deviations from GR or from the Kerr geometry, we will always take it to be small. Then 
\begin{equation}
 \sigma_X=\frac{\partial X}{\partial M}\sigma_M+\frac{\partial X}{\partial \chi}\sigma_\chi+\frac{\partial X}{\partial Q}\sigma_Q\,,
\end{equation}
where $X=(f_1,f_2,\tau_1)$. It is straightforward to solve the system of three equations above for $\sigma_M$, $\sigma_\chi$ and $\sigma_Q$; this yields
\begin{eqnarray}
 \rho\sigma_M &=& F_1(f_1,f_2,\tau_1,{\cal A}_2/{\cal A}_1) \,,\\
 \rho\sigma_\chi&=& F_2(f_1,f_2,\tau_1,{\cal A}_2/{\cal A}_1)\,,\\
 \rho\sigma_Q&=& F_3(f_1,f_2,\tau_1,{\cal A}_2/{\cal A}_1)\,,
\end{eqnarray}
where $F_i$ are, usually, cumbersome analytical functions. 
Finally, and because $Q$ is small, we can view $\sigma_Q$ as an upper bound on the quantity $Q$ itself and use it to estimate the constraint that can be imposed by a ringdown detection with a certain SNR $\rho$. This strategy was used to study constraints on the electric charge of BHs, resulting in the constraint $\frac{|Q|}{M}\lesssim 0.1 \sqrt{\frac{100}{\rho}}$~\cite{Cardoso:2016olt}.
A similar procedure can used to constraint the magnitude of the Gauss-Bonnet coupling constant, or of any other theories, as long as the QNMs are well understood in these theories~\cite{Salcedo:2016}. 
\subsubsection{Environmental effects}

The QNMs of BHs are intimately connected to the null circular geodesic, or light ring~\cite{Cardoso:2008bp,Barausse:2014tra,Barausse:2014pra,Cardoso:2016rao}.
In addition, orbits that pass inside the light ring must plunge into the BH. Indeed, the innermost stable circular orbit (ISCO)
for example, is always outside the light ring. As such, stars or matter debris composing accretion disks are not expected to populate the region close to the light ring.
In turn, this means that ``environmental effects'' (i.e., the impact of matter surrounding the BH) are expected to have a small impact on the QNMs of BHs and consequently on tests of the Kerr-hypothesis.

An exhaustive list of environmental effects, such as the impact of electric charge, accretion disks, dark matter, cosmological constant, etc are presented and quantified in Refs.~\cite{Barausse:2014tra,Barausse:2014pra}.
The results are summarized in Table~\ref{bstable}.
\begin{table}[b]
\centering
\footnotesize
\caption{\footnotesize Upper limits on the environmental corrections to the BH QNMs.
{We define $\delta_{R,I}=1-\omega_{R,I}/\omega_{R,I}^{(0)}$, where $\omega_{R,I}$ is the real (imaginary)
part of the ringdown frequency in the presence of environmental effects, whereas $\omega_{R,I}^{(0)}$ 
is the same for an isolated BH with the same total mass.
Conservative environmental reference values are $q=10^{-3},\,B=10^{8}\,{\rm Gauss},\,\rho_3^{\rm DM}=\rho_{\rm DM}/(10^3 M_\odot/{\rm pc}^3)$. We assume a Shakura-Sunyaev disk model with viscosity parameter $\alpha=0.1$
and Eddington ratio $f_{\rm Edd}=10^{-4}$ ($f_{\rm Edd}=1$) for thick and thin disks, respectively. 
The spherical and ring-like matter distributions  have  mass $\delta M \sim 10^{-3}M$.
The scaling with the parameters is shown in Ref.~\cite{Barausse:2014tra}. Taken from Ref.~\cite{Barausse:2014pra}.
}
}
\vskip 0.2cm
\begin{tabular}{c|cc}
 \hline\hline

Correction           	   	& $|\delta_R|[\%]$ 	&$|\delta_I|[\%]$   	\\
\hline
spherical near-horizon distribution
	&$0.05$	             	&$0.03$             	\\
ring at ISCO	 	       	&$0.01$	             	&$0.01$	            \\
electric charge 		               	&$10^{-5} $   	&$10^{-6}$    	\\
magnetic field           	&$10^{-8}$  	&$10^{-7}$	\\
gas accretion              	&$10^{-11} $  	&$10^{-11}$     \\
DM halos	 	       	&$10^{-21}\rho_3^{\rm DM}$ 	&$10^{-21}\rho_3^{\rm DM}$   	\\
cosmological effects 		&$10^{-32}$ 	&$10^{-32}$   	\\
\hline\hline
\end{tabular}
\vspace{0.2cm}
\label{bstable}
\end{table}
These effects represent the ultimate precision with which one can test the no-hair hypothesis, in the absence of more detailed knowledge about the matter surrounding the BH.
To summarize, no-hair tests are possible with ringdown waves from perturbed BHs, with threshold signal-to-noise ratios
shown in Fig.~\ref{fig:minimumSNR}. Third-generation detectors will be able to perform accurate tests of GR, or possibly infer details of the environment surrounding BHs.
\subsubsection{QNMs as a probe of the event horizon}
As we remarked, the BH ringdown is a dynamical probe of the underlying theory, and is tightly connected to the light ring properties.
It is not surprising therefore that ringdown probes the region very close to the event horizon~\cite{Nakamura:2016gri,Kinugawa:2016mfs,Nakamura:2016yjl}.
However, precisely because ringdown is related to the light ring, ultra-compact objects with light rings may also display similar signals~\cite{Cardoso:2016rao,Barausse:2014tra}.

\subsection{Dynamical tests with gravitational waves: inspiral}\label{sec:Ryan}
As shown by Ryan~\cite{Ryan:1995wh}, the multipole moments of a central object which is assumed to be stationary,
axisymmetric and reflection-symmetric, can be measured by looking at the motion of a test body orbiting around the
central object. Indeed, observable, gauge-invariant quantities can be expressed in terms of the multipole moments, as
defined by Geroch and Hansen. These quantities are the GW spectrum, i.e. (as mentioned in Sec.~\ref{sec:multipolesngr}) the
amount of GW energy emitted per logarithmic interval of frequency; the perihastron precession and the orbital plane
precession (``epyciclic frequencies''); the number of cycles of the GW signal emitted per logarithmic interval. For
instance, the GW spectrum is
\begin{eqnarray}
  \Delta E(f)&=&f\frac{dE_{gw}}{df}\nonumber\\
  &=&\mu\left(\frac{1}{3}v^2-\frac{1}{2}v^4+
  \frac{20}{9}\frac{S_1}{M^2}v^5+\left(-\frac{27}{8}+\frac{M_2}{M^3}\right)v^6+\dots\right)\label{deltaEmom}
\end{eqnarray}
where $f$ is the frequency of the emitted GWs (twice the orbital frequency $\nu$), $\mu$ is the mass of the test body,
and $v=(\pi Mf)^{1/3}$. Since the frequency is associated with the Killing vector of the stationary spacetime, all of
these quantities are gauge-invariant. Note that the post-Newtonian expansion~(\ref{deltaEmom}) is only accurate for
$v\ll1$; near the ISCO, $v\sim(6)^{-1/2}$ (or larger for rotating BHs) and higher-order terms should be included in the
expansion~\cite{Fujita:2012cm}.

Ryan's construction~\cite{Ryan:1995wh}, in principle, seems to be an extremely powerful phenomenological tool. GW and
electromagnetic observations can be used to measure the multipole moments of the central object, thus providing a
mapping of the spacetime. In particular, the moments can be extracted from the GW signal emitted by an extreme
mass-ratio inspiral (EMRI), i.e., a stellar-mass BH inspiralling around a supermassive BH - a target source for
space-based detectors such as eLISA~\cite{Seoane:2013qna,Aasi:2013wya}. A comparison with the moments of Kerr's
solution~(\ref{Kerrmoments}) would provide a test of the ``no-hair'' hypothesis.

In practice, the ambitious program of ``mapping the BH spacetime'' with its multipole moments turned out to be less
practical than expected. Firstly, it has been observed that multipole expansions are poorly convergent in the strong-field
region near the BH horizon, where a large number of terms is required to describe the
spacetime~\cite{Collins:2004ex}. Although the detection of an EMRI waveform by an eLISA-like detector could allow to
measure the quadrupole $Q=M_2$ with good accuracy~\cite{Ryan:1997hg,Barack:2006pq}, a deviation from Kerr spacetime can
affect the higher-order multipoles as well, which are difficult to measure independently.  Secondly, the beautiful
expression of $\Delta E$ in terms of the multipole moments~(\ref{deltaEmom}) is useless if we do not also know the
geodesic motion and the emitted GW flux, which are required to compute the gravitational
waveform~\cite{Glampedakis:2005cf}; and both geodesic motion and GW flux are very difficult to describe
for a non-Kerr spacetime with arbitrary multipole moments. We also remark that the GW flux depends on
the dynamical equations of gravity.

Several different approaches have been developed to deal with these concerns. Ryan's construction has been extended to
specific modified gravity theories, such as scalar-tensor gravity~\cite{Pappas:2014gca} and, to some extent, to
EDGB gravity~\cite{Kleihaus:2014lba}, in order to make the ``theory bias'' towards general
relativity less severe. Other authors have constructed parametrized deviations of Kerr spacetimes (``bumpy'' or
``quasi-Kerr'' BHs), studying geodesics and EMRI waveforms. In some of these solutions, the deviation only affects the
quadrupole moment~\cite{Collins:2004ex,Glampedakis:2005cf}; in others, higher-order multipoles are also
affected~\cite{manko1992generalizations,Gair:2007kr,Johannsen:2011dh,Vigeland:2009pr,Vigeland:2011ji,Cardoso:2014rha,Johannsen:2015pca}. However,
in many of these ``bumpy'' BH solutions the spacetime metric is not parametrized by multipole moments. These solutions
have also been used to devise non-dynamical tests of the ``no-hair'' hypotheses based on astrophysical, electromagnetic
observations (see Sec.~\ref{sec:emobs}).

\subsection{Tests with electromagnetic observations}\label{sec:emobs}
Non-dynamical tests of the ``no-hair'' hypothesis can also be done with astrophysical
observations of the electromagnetic emission from the surroundings of BH candidates (see
e.g.~\cite{Psaltis:2008bb,Bambi:2015kza,Johannsen:2016uoh} and references therein).
\subsubsection{Motion of stars}
A promising measurement of the multipole moments of BH candidates is the (electromagnetic) observation of stars on tight
orbits around supermassive BHs, most specially Sgr A*,
the compact object at the center of our galaxy~\cite{Will:2007pp,Merritt:2009ex,Psaltis:2015uza,Christian:2015smg}.
Although such tests are complicated by several additional factors (related to the fact that such orbits have typical
radii much larger than those involved in GW observations), progress in instrumentation makes them attractive
possibilities in the near-future.

The idea is very simple, and consists on measuring the pericenter and orbital plane precession of stars orbiting a massive BH, 
on tight and eccentric enough orbits. The precession depends on the mass of the central object (the Schwarzschild - $S$ - part of the geometry), on the spin $J$ and on the quadrupole moment $Q$ of the BH and can therefore
be inverted to estimate each of these quantities, testing the no-hair hypothesis. 

Define,
\beq
A_S&=&\frac{6\pi}{c^2}\frac{GM}{a(1-e^2)}\,,\\
A_J&=&\frac{4\pi\,j}{c^3}\left[\frac{GM}{a(1-e^2)}\right]^{3/2}\,,\\
A_Q&=&\frac{3\pi\,j^2}{c^4}\left[\frac{GM}{a(1-e^2)}\right]^{2}\,,
\eeq
with $M$ the central, massive object, $j$ its dimensionless spin, $e$ the star's eccentricity and $a$ its semi-major axis.
In the orbit-averaged approximation, and assuming that the orbiting stars are much lighter than the central object,
stars experience an advance of the orbital periapse given by (to lowest post-Newtonian order)
\be
{\bf \delta \varpi}=A_S-2A_J-\frac{1}{2}A_Q(1-3\cos^2i)\,,
\ee
per orbit. In addition the orbital plane also precesses due to coupling between the orbital angular momentum and the central object spin $J$.
The nodal precession is 
\be
\delta\Omega=A_J-A_Q\cos i\,,
\ee
with $i$ the orbital inclination.

It can be seen that the Schwarzschild contribution exceeds the spin and quadrupole terms for in-plane precession, for most of the parameter space.
Fortunately, the orbital plane precession depends only on the spin and quadrupole contributions. However, to test the no-hair hypothesis,
one must determine five parameters (the BH mass, quadrupole moment and the three spin components), thus the orbits of two stars are required~\cite{Will:2007pp}. 

\begin{figure}
\begin{center}
\begin{tabular}{c}
\includegraphics[width=0.6\textwidth]{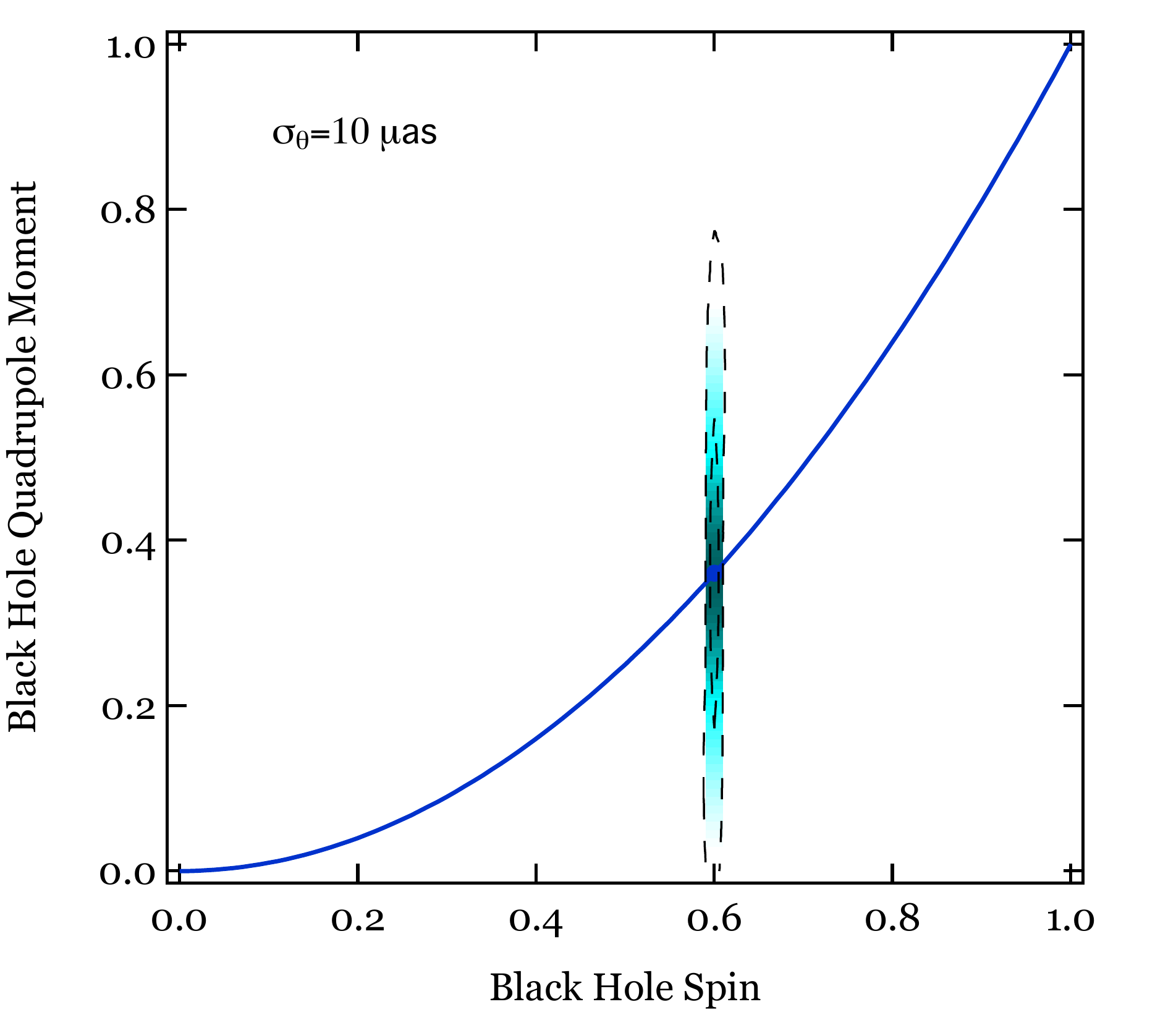} \\
\end{tabular}
\end{center}
\caption{\label{fig:stars_nohair} The posterior likelihood of measuring the spin and quadrupole
  moment of Sgr A$^*$ by tracing the orbits of two stars with GRAVITY,
  assuming an astrometric precision of $10~\mu$as. The dashed curves show the 68\% and
  95\% confidence limits, while the solid curve shows the expected
  relation between these two quantities in the Kerr metric.  The assumed (dimensionless)
  spin and quadrupole moment are $j=0.6$, $Q/M^3=0.36$). The two stars are assumed to have
  orbital separations equal to 800\,$M$ and 1000\,$M$ and
  eccentricities of 0.9 and 0.8, respectively. Even at these
  relatively small orbital separations, tracing the orbits of stars
  primarily measures the spin of the BH, unless a very high level of astrometric precision is achieved. 
	From Ref.~\cite{Psaltis:2015uza}.
}
\end{figure}
%
In addition, there are a number of effects that complicate the determination of the BH's multipole moments, such as~\cite{Will:2007pp,Merritt:2009ex,Psaltis:2015uza}

\noindent {\bf i.} The contribution of other orbiting stars to the mass, spin and quadrupole moment of the central object.
This contribution can be estimated once a stellar density distribution is known or prescribed. Assuming a number density of stars 
\be
n(a,\theta)=N_0/a_0^3(a/a_0)^{-2}n_{\theta}(\theta)\,,
\ee
one gets for the relative contributions to the mass, spin and quadrupole moment~\cite{Merritt:2009ex,Psaltis:2015uza},
\beq
\frac{\delta M}{M}&=&4.8\times10^{-8}M_6\left(\frac{a_0}{1\,{\rm pc}}\right)^{-1}a_1\,,\\
j\frac{\delta J}{J}&=&3\times10^{-8}M_6\left(\frac{a_0}{1\,{\rm pc}}\right)^{-1}a_1^{3/2}\,,\\
Q/M^3\frac{\delta Q}{Q}&=&4\times10^{-10}M_6\frac{\tilde{n}_\theta}{0.1}\left(\frac{a_0}{1\,{\rm pc}}\right)^{-1}a_1^3\,.
\eeq
Here, $a_1\equiv\frac{a}{M}$, $M_6\equiv \frac{M_*}{10^6M_{\odot}}$, $M_*$ being the total mass of stars inside a characteristic orbital separation $a_0$, and 
\be
\tilde{n}_{\theta}=\int_{-1}^{1}(3\cos^2\theta-1)n_{\theta}(\theta)d\cos\theta\,,
\ee
characterizes the angular distribution of stars. The strongest constraint here comes from the quadrupole corrections, which depend sensitively on the semi-major axis $a$.
For the corrections to be under control, $a\lesssim 1000 M$.

\noindent {\bf ii.} Perturbations to the orbit of the stars being measured. These effects include

\noindent {\bf iia.} Decoherence of the orbit due to Newtonian interactions with other stars, on a timescale
\be
t_N=12.6\times 10^8 M_6^{-1/2}\left(\frac{a_0}{1\,{\rm pc}}\right)^{-1/2} a_1\left(\frac{m_*}{M_{\odot}}\right)^{-1/2}\left(\frac{M}{4.3\times 10^6M_{\odot}}\right)^{3/2}\,{\rm sec},
\ee
where $m_*$ is the average mass of a star within the stellar cluster.

\noindent {\bf iib.} Decoherence is the dominant perturbing effect, but in hydrodynamic interactions with the accretion flow (drag), stellar winds and tidal effects on the stars being measured 
introduce additional sources of error. These are quantified in Ref.~\cite{Psaltis:2015uza}.

In Ref.~\cite{Psaltis:2015uza} the authors have studied how one can constrain the spin and quadrupole moment of Sgr A$^*$, 
by measuring the motion of two stars on eccentric orbits. The results are shown in Fig.~\ref{fig:stars_nohair}, assuming an astrometric precision of $10~\mu$as.
It turns out that, even at the small orbital separations used in their study, tracing the orbits of stars primarily measures the spin of the BH. 

We note, however, that even such large errors on the BH quadrupole moment already impose some constraints on some hairy BH solutions 
(see, e.g, Ref.~\cite{Herdeiro:2014goa}, where some of the BH solutions have extremely large quadrupole moments).

\subsubsection{Motion of pulsars}

It was pointed out some time ago that the observation of a {\it single} pulsar in orbit around a very compact object
might still allow for tests of the Kerr hypothesis~\cite{Wex:1998wt}. The basic idea rests, of course, on data available through pulsar timing.
The strategy and sources of error are described in detail in Refs.~\cite{Liu:2011ae,Psaltis:2015uza}.
The mass of the central object (and the inclination of the orbital plane) can be determined from the precession of the periastron or via Shapiro delay. 
The Lense-Thirring precession, along with a measurement of the precession of periastron, the projected semi-major axis and their time derivatives,
allow for the determination of all three spin components. Finally, Roemer delay can be used to estimate the quadrupole moment~\cite{Liu:2011ae}.

Tests of the Kerr hypothesis using single pulsars requires high-eccentricity pulsar and sub-year orbital periods, but is a promising tool for the near-future~\cite{Liu:2011ae,Psaltis:2015uza}.

\subsubsection{Accretion disks\label{sec:accretion}}
Matter accretion onto BHs is one of the most luminous phenomena in the universe: the efficiency for converting rest mass
into radiation in this process can be as large as $\sim10\%$, much larger than, i.e., thermonuclear processes.  The
radiation emitted by accreting matter at few gravitational radii from the horizon, can be a promising probe of the BH
strong-field region.

The matter surrounding a BH forms an accretion disk~\cite{Novikov:1973kta,page1974disk}, in which each element
approximately moves in circular, Keplerian orbits.  As the matter element loses angular momentum, it moves inward to a
different circular orbit, until it reaches the inner edge of the disk, corresponding to the
ISCO of the BH spacetime. Then, since there is no stable orbit inside the ISCO, the matter dynamically falls
into the BH. As the ISCO represents a transition point in the physics of the accretion disk, it should be possible to
extract its location, $r_{ISCO}$, from the electromagnetic signal emitted (mostly in the X-ray band) by the accreting
matter.

As noticed in~\cite{Shibata:1998xw}, $r_{ISCO}$ (or, equivalently, the ISCO angular velocity $\Omega_{ISCO}$) is
characterized by the first multipole moments of the central object. In particular, the contribution of the quadrupole
moment to $r_{ISCO}$ is comparable to that of the BH spin. Therefore, a measurement of $r_{ISCO}$ or $\Omega_{ISCO}$,
either for a stellar-mass BH or for a supermassive BH, would be a (non-dynamical) test of the ``no-hair'' hypothesis.
In practice, since the physics of accretion disks is very complex, is not easy to extract $r_{ISCO}$ with good accuracy
from the electromagnetic signal from accreting BHs. The two main approaches to extract information on the BH spacetime
from the X-ray emission of the accretion disk are the analysis of the iron $K\alpha$ line~\cite{fabian1989x}, and the
study of the thermal component of the spectrum using the so-called continuum-fitting method ~\cite{Li:2004aq}. Presently, they allow to measure (with some confidence)
the value of BH spins~\cite{McClintock:2013vwa,Reynolds:2013qqa}, but the constraints on the BH spacetime (see
e.g.~\cite{Bambi:2015kza,Johannsen:2016uoh} and references therein) are still weak.

The iron $K\alpha$ line is the brightest component of the X-ray emission from the accretion disk of supermassive BHs, and one of the brightest components in the case of stellar-mass BHs. 
It is broadened and skewed due to (special and general) relativistic effects and Doppler effect, which
determine a characteristic shape. An analysis of this shape (assuming that the spacetime is described by the Kerr
metric) allows to measure the BH spin and the inclination of the accretion disk, even if the BH mass is
unknown~\cite{Reynolds:2013qqa}. In order to test the BH spacetime, the theoretical model of the line shape has been
extended to parametrized deviations of the Kerr spacetime (``quasi-Kerr'' or ``bumpy'' BHs, see
Sec.~\ref{sec:Ryan}); a comparison of these models with
future, more accurate observations of the iron $K\alpha$ line can allow to set bounds to the parameters characterizing
these solutions, and then to test the ``no-hair''
hypothesis~\cite{Johannsen:2010xs,Johannsen:2012ng,Bambi:2012at,Jiang:2014loa,Jiang:2015dla,Johannsen:2015rca,Moore:2015bxa,Hoormann:2016dhy}.
The main limitation of this approach is that current theoretical models of the iron line are too simple to prevent
systematic effects, which may dominate possible deviations from Kerr spacetime~\cite{Bambi:2015kza}.

The spectrum emitted by accretion disks of stellar-mass BHs has a strong thermal component, which can be analysed with
the continuum-fitting method, i.e. by comparison with a theoretical model, in order to measure $r_{ISCO}$.  In this
model the standard, Novikov-Thorne description of of the accretion disk~\cite{Novikov:1973kta,page1974disk} is assumed:
the disk is geometrically thin, optically thick, orthogonal to the BH spin, with matter moving in circular geodesics,
and inner edge at $r_{ISCO}$ (this description is believed to be accurate for a subset of the actual BH accretion
disks). The background spacetime is assumed to be described by the Kerr metric. If the mass, distance and disk
inclination are known from independent measurements, this approach allows to determine $r_{ISCO}$ and then to measure
the BH spin~\cite{McClintock:2013vwa}. This model has been generalized to ``quasi-Kerr'' or ``bumpy'' BH spacetimes, in
order to set bounds, comparing the model with observational data, to the deviations from the Kerr metric, and then to
test the ``no-hair''
hypothesis~\cite{Johannsen:2010xs,Bambi:2011jq,Bambi:2012tg,Kong:2014wha,Bambi:2014sfa,Johannsen:2015rca,Moore:2015bxa,Hoormann:2016dhy}. While
the continuum-fitting method can provide, assuming the Kerr background, a reliable measurement of the BH spin for a
significant fraction of the stellar-mass BHs, it is much less effective in constraining deviations from the Kerr
metric. Indeed, the parameters characterizing the deviation are typically degenerate with
$r_{ISCO}$~\cite{Bambi:2015kza,Johannsen:2016uoh}. This degeneracy is partially removed when the ISCO is very close to
the horizon, i.e. for near-extremal BHs.
\subsubsection{Black hole ``shadows''}
%
\begin{figure}
\begin{center}
\begin{tabular}{c}
\includegraphics[width=0.6\textwidth]{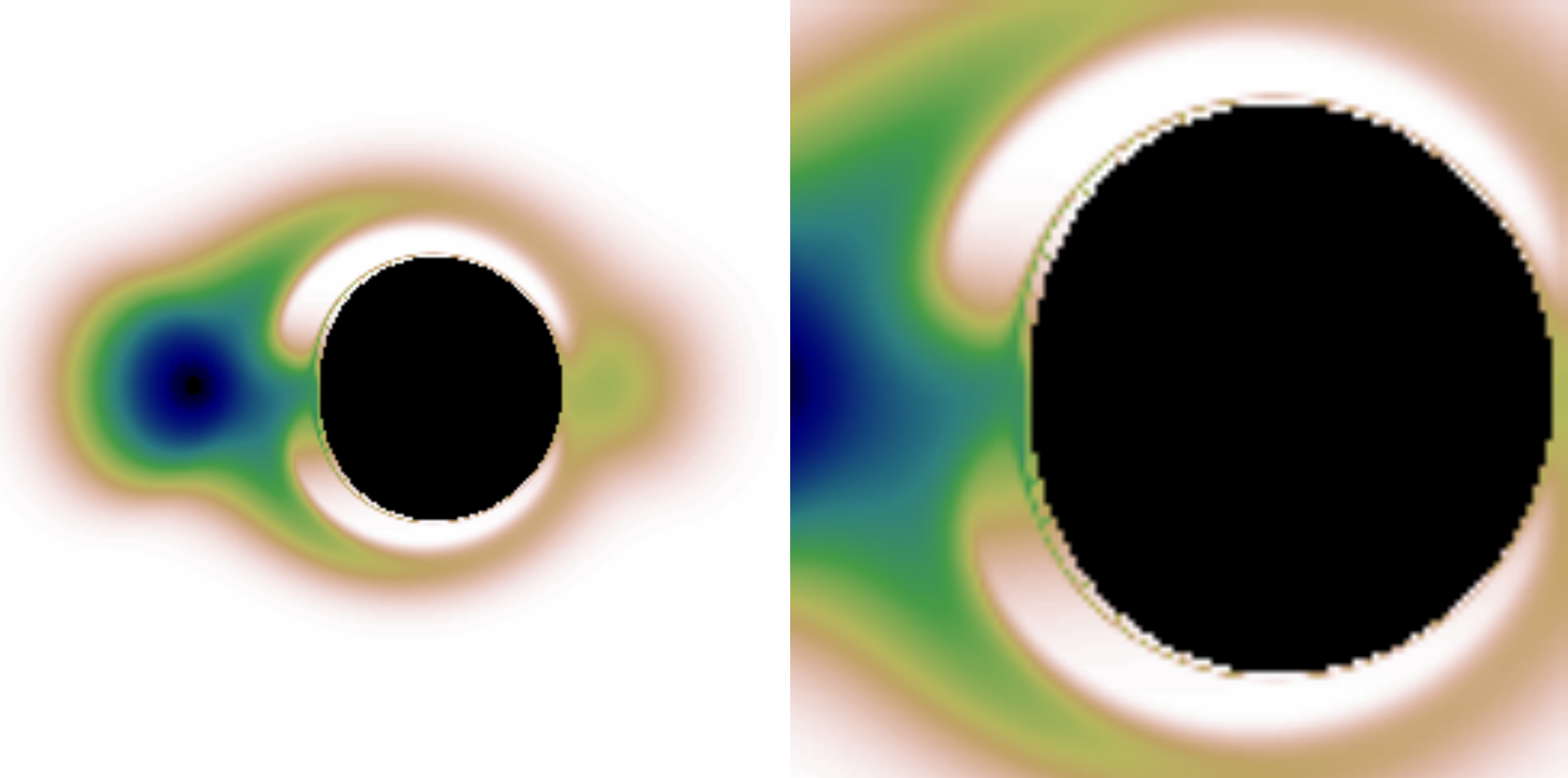} \\
\end{tabular}
\end{center}
\caption{\label{fig:shadow_kerr} Image at $\lambda=1.3$~mm of an accretion torus surrounding a Kerr BH. 
The Kerr BH is spinning with $a/M=0.99$, and the torus is being observed at an angle $i=85^o$.
	The torus has an inner radius of $4.2M$, temperature $5.3\times 10^{10}$ K and a polytropic index $5/3$.
	For further details see Ref.~\cite{Vincent:2015xta}.
The color bar indicates the cgs value of specific intensity.
	The directions on the observer's sky that asymptotically approach the event horizon when ray tracing
	backwards in time are marked in black color. The black area at the center of the image is
	the BH shadow. Its exterior limit nearly coincides with the light ring. The right panel shows a zoom on the central region.
	From Ref.~\cite{Vincent:2015xta}.
}
\end{figure}
As discussed in Sec.~\ref{sec:Ryan}, when matter is moving very close to a BH (or any compact object) a multipolar decomposition of the gravitational field is not particularly useful since all, or a substantial number of, multipoles contribute to the gravitational potential and to the motion of matter. However, the spacetime around compact objects possesses unique features -- such as innermost stable circular orbits, light rings (unstable null geodesics), etc~\cite{Bardeen:1972fi} -- that might be used as smoking guns of the BH-nature of the object and even of GR. In particular, the null geodesics carry information about the effective size of BHs, since in essence any particle or light ray penetrating the light ring will never reach asymptotic observers. Thus, BHs create ``shadows'' of matter around them~\cite{1975ApJ...202..788C,1979AA....75..228L,Falcke:1999pj}.
The exact shape and appearance of BHs depends on the source illuminating them, but a crucial ingredient determining the optical appearance is the rotation rate, which determines how close to the horizon the co-rotating light ring is, and how far away from the horizon the counter-rotating light ring is. The rotation rate also determines how tight the accretion disk can bind to the BH, and how much the gravitational and Doppler shift emitted from the disk will be.
An example of the optical appearance of BHs is shown in Fig.~\ref{fig:shadow_kerr}, for a Kerr BH rotating with $a/M=0.9$ and being illuminated by an accretion torus~\cite{Vincent:2015xta}.

\begin{figure}
\begin{center}
\begin{tabular}{ccc}
\includegraphics[width=0.35\textwidth]{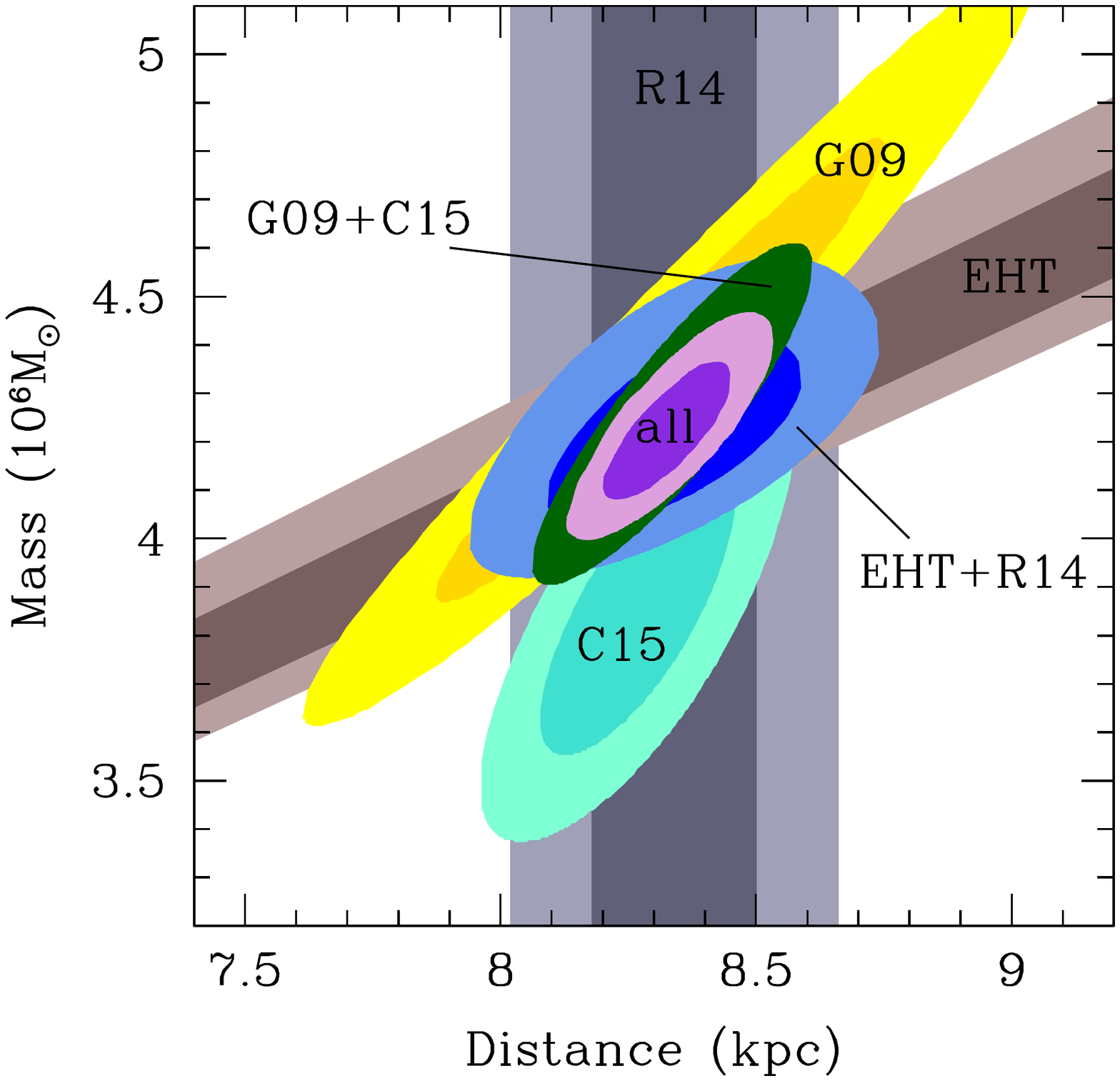}&
\includegraphics[width=0.35\textwidth]{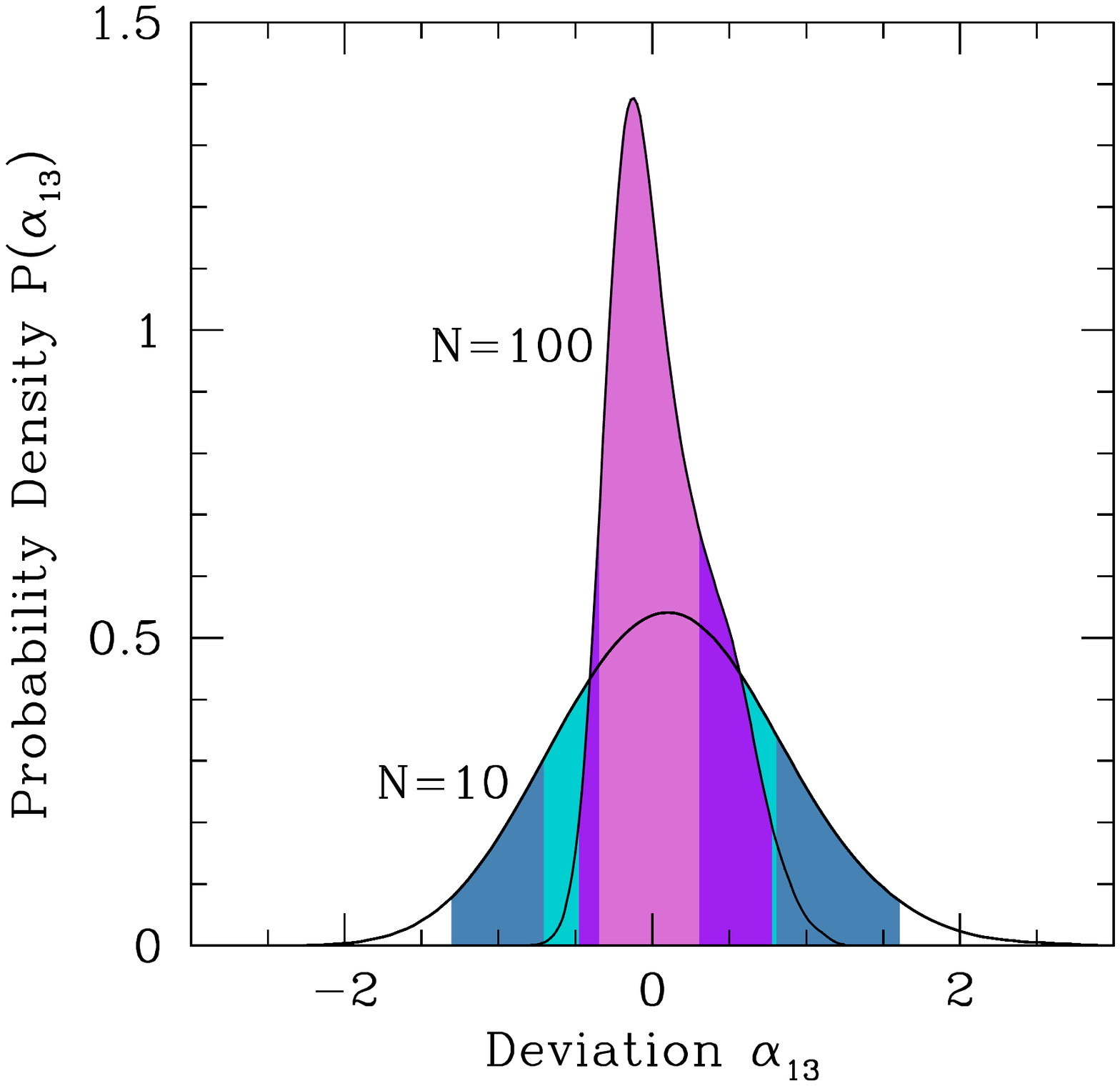}&
\includegraphics[width=0.35\textwidth]{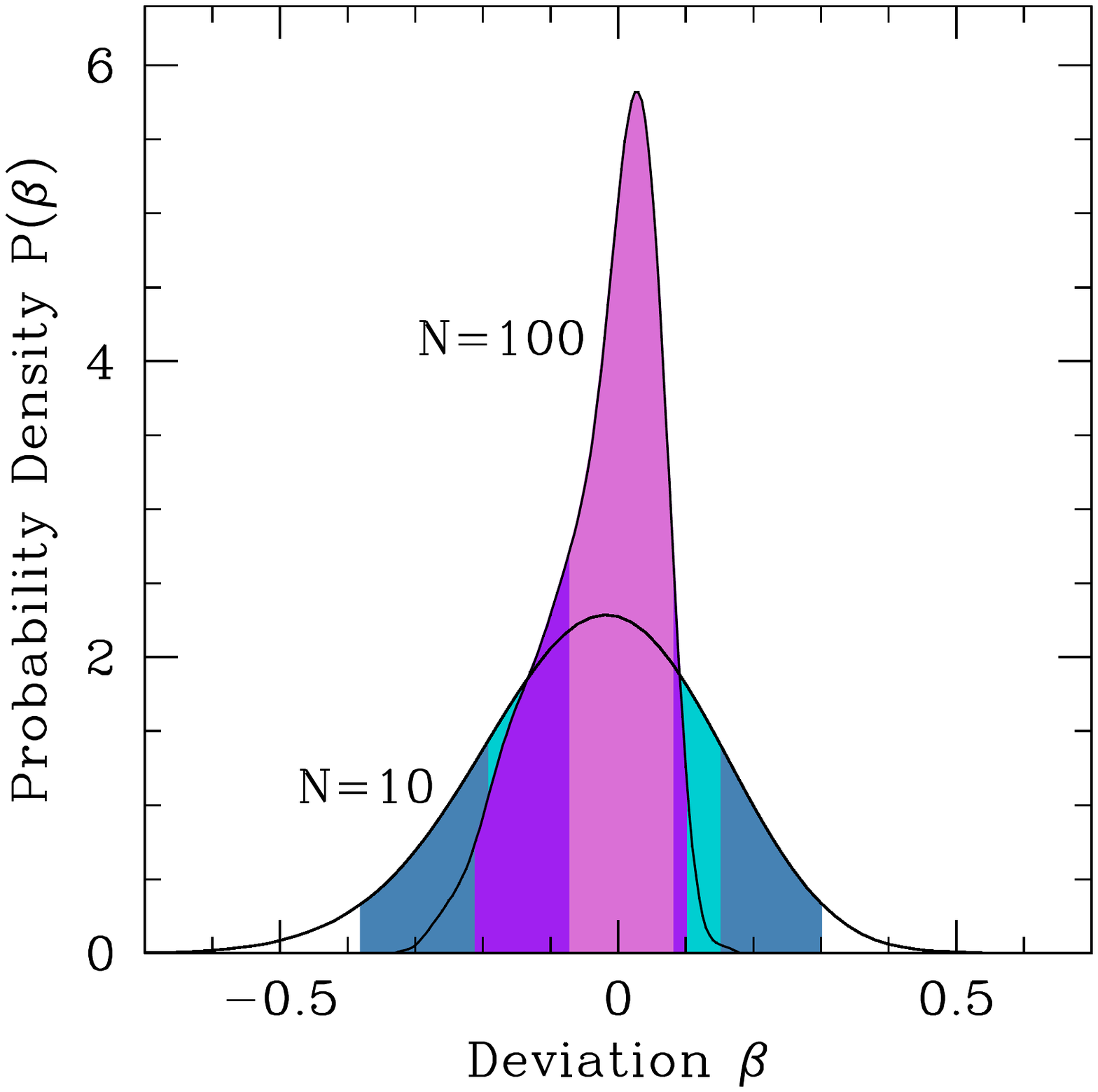}
\end{tabular}
\end{center}
\caption{Left panel: $1\sigma$ and $2\sigma$ confidence contours of the probability density of the mass and distance of Sgr~A* for existing measurements (S-stars, ``G09''~\cite{Gillessen:2008qv}; masers, ``R14''~\cite{Reid:2014boa}; star cluster, ``C15''~\cite{2015MNRAS.447..948C}), a simulated measurement of the shadow size of Sgr~A* for $N=10$ observations with a seven-station EHT array (``EHT''), and several combinations thereof. The simulated EHT measurement improves the other constraints on the mass and distance significantly. Center and right panels: Simulated $1\sigma$ and $2\sigma$ confidence contours of the probability density of the deviation parameters $\alpha_{13}$ and $\beta$, respectively, corresponding to $N=10$ and $N=100$ EHT observations, each marginalized over the mass and distance using the combination of all data sets (``all'') in the $N=10$ case and of simulated stellar-orbit observations from a 30m-class telescope~\cite{Weinberg:2004nj} in the $N=100$ case. From Ref.~\cite{Johannsen:2015hib}.
}
\label{fig:constraints}
\end{figure}
Observations of BH shadows as a tool to probe the geometry close to the event horizon became possible
with the advent of powerful instruments such as the Event Horizon Telescope~\cite{Doeleman:2008qh}.
The main obstacles to performing tests of GR with observations of BH shadows are
(a) the large number of parameters that describe the shadow, including the inclination angle of the object,
the mass and angular momentum of the BH, and all the details of the accretion mechanism (the illuminating source);
(b) the lack of a robust parametrization of deviations from the Kerr geometry, in the strong-field limit. 

Recently, a proof-of-principle describing tests of the Kerr hypothesis with observations of BH shadows was put
forward~\cite{Johannsen:2015hib}. 
As a background, the following metric (belonging to the class of ``quasi-Kerr'' spacetimes introduced
in~\cite{Johannsen:2015pca}) was proposed,
\begin{eqnarray}
ds^2&=&-\frac{\Sigma\left(r^2-2Mr+a^2\cos^2\theta\right)}{\left(A_1(r^2+a^2)-a^2\sin^2\theta\right)^2}dt^2 -\frac{2a\sin^2\theta\Sigma\left(A_1(r^2+a^2)-\tilde{\Delta}\right)}{\left(A_1(r^2+a^2)-a^2\sin^2\theta\right)^2}dtd\phi\nonumber\\ 
&+&\frac{\Sigma}{\tilde{\Delta}}dr^2+\Sigma d\theta^2+\left[\frac{A_1^2(r^2+a^2)^2-a^2\tilde{\Delta}\sin^2\theta}{\left(A_1(r^2+a^2)-a^2\sin^2\theta\right)^2}\right]\Sigma\sin^2\theta d\phi^2\,, \label{metricKerr_mod}
\end{eqnarray}
where $\tilde{\Delta}=\Delta+\beta M^2$ and $A_1=1+\alpha_{13}M^3/r^3$. 
This geometry is parametrized by the BH mass $M$, angular momentum $J=Ma$ and the constants $\alpha_{13},\,\beta$.
We should point out that this particular form of the metric is rather ad-hoc and does not describe any known theory, but
that in any case the results are to be taken only as proof-of-principle.

The shadow resulting from the metric (\ref{metricKerr_mod}) was compared to simulations of observations by the Event Horizon Telescope of the BH at the center of our galaxy, Sgr~A*. The result is summarized in Fig.~\ref{fig:constraints}. The study uses a reconstructed image of Sgr~A* based on a simulated image of a radiatively-inefficient accretion flow.
Within the working hypothesis, the prospects for constraining the unknown parameters $\beta,\,\alpha_{13}$ are good: with 100 observations, the parameters can be determined to be
\beq
\alpha_{13}&=&0.13^{+0.43}_{-0.21} {}^{+0.9}_{-0.34}\\
\beta &=&0.03^{+0.05}_{-0.10} {}^{+0.07}_{-0.24}\,.
\eeq
%
The uncertainties associated with these parameters are such that they are effectively of order ${\cal O}(1)$, precisely the order at which they enter in the metric\footnote[3]{Note also that, when trying to translate such constraints to specific modified theories, care has to be exercised: in EDGB theory, for example, the natural parameter that appears in the equations is $\alpha'/M^2$ where $\alpha'$
is the theory's coupling constant, with dimensions of mass squared~\cite{Pani:2009wy,Berti:2015itd}. Effectively then, one is constraining $\alpha'/M^2$ which yields {\it poor}
constraints on the coupling constant itself for supermassive BHs.}.

Hairy BHs arising naturally from simple theories, such as the minimally coupled scalar
theory~(\ref{minimally_coupled})~\cite{Herdeiro:2014goa,Herdeiro:2015waa}, will also give rise to shadows which can, in
principle, be discriminated from those of Kerr BHs~\cite{Cunha:2015yba}.

It should be noted that shadow observations can also, in principle, be used to test the BH-nature of compact objects.
One could expect that objects without a horizon, and presumably with a surface, will feature a bright surface, allowing immediately to discriminate between BHs and, say, gravastars. It was pointed out in Ref.\cite{Vincent:2015xta} that the main factor determining the shadow is the light ring. As such, horizonless compact objects with a light ring can mimick Kerr BHs very efficintly, {\it provided}
the illuminating source is not accreted towards the center of the object.

%
%

\section{Conclusions}
It is by now well-known that the Kerr family is {\it not} the most general solution of Einstein's field equations
in the presence of reasonable forms of matter. However, there are good reasons to believe that
BHs which do not belong to the Kerr family are either dynamically unstable or do not form out of realistic collapse scenarios.

The no-hair or Kerr hypothesis is therefore a cherished belief, which upcoming experiments can test, either in the GW or electromagnetic band. Some of these measurements will be more sensitive to the null geodesic around compact objects, than to the presence of the event horizon itself~\cite{Cardoso:2016rao}... but given that even light rings are a unique feature of relativistic theories of gravity, 
these are all exciting years ahead.


\vspace{1cm}
\noindent
{\bf Acknowledgments.}
%
We thank the anonymous referee for useful suggestions, which contributed to improve the quality of the manuscript.
We are indebted to Jo\~ao Costa, Carlos Herdeiro, Chris Moore, Thomas Sotiriou, Norbert Wex, and Nicolas Yunes for useful comments on the manuscript.
V. C thanks the Physics Department of the University of Rome ``La Sapienza'' for hospitality while this work was being completed.
V.C. acknowledges financial support provided under the European Union's H2020 ERC Consolidator Grant ``Matter and strong-field gravity: New frontiers in Einstein's theory'' grant agreement no. MaGRaTh--646597,
and FCT for Sabbatical Fellowship nr. SFRH/BSAB/105955/2014.
Research at Perimeter Institute is supported by the Government of Canada through Industry Canada and by the Province of Ontario through the Ministry of Economic Development $\&$
Innovation.
This work was supported by the H2020-MSCA-RISE-2015 Grant No. StronGrHEP-690904.
%
\vskip 1cm

\addcontentsline{toc}{section}{Appendix: Multipole moments in Newtonian gravity and in general relativity}
\section*{Multipole moments in Newtonian gravity and in general relativity}
In this Appendix we describe the multipole expansion framework, discussed in Section~\ref{sec:multipoles}, both in
Newtonian gravity and in GR. For a more detailed discussion, we refer the reader
to~\cite{jackson1999classical,poisson2014gravity} (Newtonian gravity)
and~\cite{Geroch:1970cd,Hansen:1974zz,Thorne:1980ru,quevedo1990multipole} (GR).
\subsection*{Multipole moments in Newtonian gravity}
\addcontentsline{toc}{subsection}{A1: Multipole moments in Newtonian gravity}

The Newtonian gravitational potential in the exterior of a massive body with density $\rho(t,\vec x)$ is the solution of
Poisson's equation~$\nabla^2\Phi=4\pi G\rho$ in vacuum:
\begin{equation}
\Phi(t,\vec x)=-G\int\frac{\rho(t,{\vec x}')}{|\vec x-{\vec x}'|}d^3x'\,.
\end{equation}
With a Taylor expansion of $1/|\vec{x}-{\vec x}'|$ around ${\vec x}'=\vec0$ and a spherical harmonic decomposition, this
solution can be written as a series in $1/r$ (where $r=|{\vec x}|$), which is called {\it multipolar expansion of the
  potential}:
\begin{equation}
\Phi(t,\vec x)=-G\sum_{lm}\frac{1}{r^{l+1}}\frac{4\pi}{2l+1}I_{lm}(t)Y_{lm}(\theta,\phi)\,,\label{newtmultexp}
\end{equation}
where $Y_{lm}$ are the usual spherical harmonics, and
\begin{equation}
I_{lm}(t)=\int\rho(t,{\vec x})r^lY_{lm}(\theta,\phi)d^3x\label{multipolessh}
\end{equation}
are the {\it multipole moments} of the potential.

This expansion can also be expressed in terms of symmetric-trace-free (STF) tensors~\cite{Thorne:1980ru,jackson1999classical}:
\begin{equation}
\Phi(t,\vec x)=-G\sum_{l=0}^\infty\frac{1}{r^{l+1}}\frac{(2l-1)!!}{l!}I^{<i_1\cdots i_l>}n^{i_1}\cdots n^{i_l}\label{newtmultexpstf}\,,
\end{equation}
where $(2l-1)!!=(2l-1)(2l-3)(2l-5)\cdots1$, $n^i=x^i/r$ (in polar coordinates,
$n^i=(\sin\theta\cos\phi,\sin\theta\sin\phi,\cos\theta)$), the brackets $<\cdots>$ denote the {\it symmetric and
  trace-free} part of a tensor,
and $I^{<i_1\cdots i_l>}(t)=\int \rho(t,{\vec x}')(x^{i_1}\cdots x^{i_l}-\hbox{trace parts})d^3x$ are the multipole
moments. For a given value of $l$, the components of the STF tensor $n^{<i_1}\cdots n^{i_l>}$ are combinations of the
spherical harmonics $Y_{lm}$ with $m=-l,\dots,l$, and $I^{<i_1\cdots i_l>}$ are combinations of the
multipoles~(\ref{multipolessh}) $I_{lm}$. The first terms of this expansion are
\begin{equation}
\Phi=-\frac{GM}{r}-\frac{3G}{2}I^{<ij>}x^ix^j
\end{equation}
where $I=M$ is the mass, $I^i=\int\rho x^id^3x=0$ in the center-of-mass frame,
$I^{<ij>}=\int\rho(x^ix^i-\delta^{ij}r^2/3)d^3x$ is the quadrupole moment\footnote[4]{Some authors use a different
  normalization for the STF multipoles, i.e. $Q^{<i_1\cdots i_l>}=(2l-1)!!I^{<i_1\cdots i_l>}$. With this notation the
  quadrupole tensor is $Q^{<ij>}=\int\rho(3x^ix^i-\delta^{ij}r^2)d^3x$ as, e.g., in~\cite{jackson1999classical}.}.

In most cases, one considers the multipole expansion of a {\it stationary, axisymmetric} body with symmetry axis
${\hat k}=(0,0,1)$.  Then, the only non-vanishing moments are $I_{l0}$ and, defining
\begin{equation}
M_l\equiv\sqrt{\frac{4\pi}{2l+1}}I_{l0}=\int\rho({\vec x})r^lP_l(\cos\theta)d^3x\,,
\end{equation}
the expansion~(\ref{newtmultexp}) reduces to Eq.~(\ref{multnewtaxis}):
\begin{equation}
\Phi(\vec x)=-G\sum_{l=0}^\infty\frac{M_l}{r^{l+1}}P_l(\cos\theta)\label{multnewtaxis2}\,,
\end{equation}
where, as discussed in Section~\ref{sec:multipoles}, $M_0=M$, $M_1=0$ in the center-of-mass frame, $M_2=Q$ quadrupole
moment. With the further assumption that the body (and then, the gravitational potential) is {\it reflection-symmetric
  across the equatorial plane}, i.e., symmetric for $\theta\rightarrow\pi-\theta$, then $M_{2l+1}=0$: the only
non-vanishing multipoles are those with even values of $l$.

In STF notation, axi-symmetry with respect to ${\hat k}$ implies that $I^{<i_1\cdots i_l>}\propto k^{<i_1}\cdots k^{i_l>}$.
Using normalization properties of STF tensors~\cite{poisson2014gravity}, it can be shown that
\begin{equation}
  I^{<i_1\cdots i_l>}=M_lk^{<i_1}\cdots k^{i_l>}\,.
\end{equation}
For instance, the quadrupole STF tensor is $I^{<ij>}=M_2(k^ik^j-\delta^{ij}/3)=Q\,{\rm diag}(-1/3,-1/3,2/3)$.

\subsection*{Multipole moments in general relativity: the Geroch-Hansen construction}\label{app_subs_gh}
\addcontentsline{toc}{subsection}{A2: Multipole moments in general relativity: the Geroch-Hansen construction}

Geroch and Hansen developed a formalism to define and compute the multipole moments of stationary, asymptotically flat
solutions of Einstein's equations in vacuum~\cite{Geroch:1970cd,Hansen:1974zz}. Their approach has then been generalized
to non-vacuum spacetimes, including the electromagnetic field~\cite{Sotiriou:2004ud} (see
also~\cite{hoenselaers1990multipole}) and a scalar field~\cite{Pappas:2014gca}. The latter work can also be seen as a
generalization to modified gravity theories: the multipole structure with a scalar field, derived
in~\cite{Pappas:2014gca}, also applies to vacuum spacetimes in Bergmann-Wagoner scalar-tensor
theories~\cite{Bergmann:1968ve,Wagoner:1970vr}. Other scalar-tensor theories with additional terms falling off rapidly
enough, such as EDGB theory~\cite{Kanti:1995vq}, enjoy the same multipole structure, as can be
seen by comparing the derivation of the quadrupole moment in EDGB BHs~\cite{Kleihaus:2014lba}
with the computation of multipole moments in Bergmann-Wagoner theories~\cite{Pappas:2014gca}.

In the Geroch-Hansen construction, multipole moments are introduced as tensors at infinity, generated by a set of
potentials. The asymptotic behaviour is defined without introducing a specific coordinate frame, through an asymptotic
completion of spacetime in which, after a conformal rescaling of the metric, the spacetime is extended to include the
``infinite point'' $\Lambda$. To understand this procedure, let us consider for simplicity the Euclidean flat space in
Newtonian theory, with metric $ds^2=\delta^{ij}dx^idx^j$. The multipole expansion of a potential $\Phi(\vec x)$ is
\begin{equation}
  \Phi=\sum_{l\ge0}\frac{1}{l!}\frac{1}{r^{l+1}}Q^{<i_1\cdots i_{l}>}x^{i_1}\cdots x^{i_l}=
  \frac{Q}{r}+\frac{Q^i}{r^3}x^i+\frac{1}{2}\frac{Q^{<ij>}}{r^5}x^ix^j+\dots
\end{equation}
Changing coordinates to ${\bar x}^i=x^i/r^2$, the metric reads $ds^2=(\delta^{ij}d{\bar x}^id{\bar x}^j)/{\bar r}^4$.
With a conformal transformation the metric becomes ${\overline{ds}}^2=\Omega^2ds^2=\delta^{ij}d{\bar x}^id{\bar x}^j$
(where $\Omega=\bar r^2$),
regular at the
infinite point $\Lambda$ ($\bar r=0$). In the new coordinate frame, the conformally rescaled field is
\begin{equation}
  \tilde\Phi(\bar x^i)=\Omega^{-1/2}\Phi=Q+Q^i{\bar x}^i+\frac{1}{2}Q^{<ij>}{\bar x}^i{\bar x}^j+\frac{1}{6}Q^{<ijk>}
  {\bar x}^i{\bar x}^j{\bar x}^k+\dots
\end{equation}
The STF derivatives of $\tilde\Phi$ evaluated at the infinite point $\Lambda$ yield the multipole moments:
$\tilde\Phi_{|\Lambda}=Q$, $\tilde\Phi_{,i|\Lambda}=Q^i$, $\tilde\Phi_{,<ij>|\Lambda}=Q^{<ij>}$,
$\tilde\Phi_{,<ijk>|\Lambda}=Q^{<ijk>}$, and so on. 

In GR the procedure is the same, with some differences:
\begin{itemize}
\item The $3$-space is the space of orbits of the timelike Killing vector (i.e., choosing the time variable
  corresponding to the Killing vector, any $t=const.$ submanifold). This is a curved space, with its own covariant
  derivative $D_i$ ($\bar D_i$ after the conformal transformation).
\item The conformal transformation is not given explicitly; it is only required that the conformally-rescaled metric is
  smooth at the infinite point $\Lambda$, and that the conformal factor and its first derivative (but not the second)
  vanish in $\Lambda$, i.e. that $\Omega$ falls off ``like $1/r^2$''. Actually, this requirement is the definition of
  ``asymptotic flatness'' of the spacetime.
\item The derivatives on the fields are covariant derivatives $\bar D_i$. Moreover, the requirement that the moments do
  not depend on the choice of $\Omega$ determines extra terms involving the Ricci tensor (of the $3$-space) in the
  definition of the moments. For instance, $Q^{<ij>}=({\bar D}_{<i}{\bar D}_{j>}\Phi-1/2{\cal R}_{<ij>}\Phi)_{|\Lambda}$.
\end{itemize}
In this way, the asymptotic expansion of the potentials can be constructed in a coordinate-independent way; the
multipoles are defined as tensors in $\Lambda$.

The next step is to define the potentials corresponding to the multipoles of the four-dimensional spacetime. Given the
Killing vector field $\xi^\mu$, the mass potential and the angular potential are $\Phi_M=(\lambda^2+\omega^2-1)/(4\lambda)$ and
$\Phi_J=\omega/(2\lambda)$, where 
\begin{eqnarray}
\lambda&=&-\xi^\mu\xi_\mu\nonumber\\
\omega_\alpha&=&\epsilon_{\alpha\beta\gamma\delta}\xi^\beta\xi^{\delta;\gamma}=\omega_{,\alpha}
\end{eqnarray}
(the fact that $\omega_\alpha$ is a gradient is a consequence of Einstein's field equations).  The multipole expansion
of these potentials, defined with the procedure outlined above, yields the mass multipoles $M^{<i_1\cdots i_l>}$ and the
current multipoles $S^{<i_1\cdots i_l>}$. In the weak-field limit, the Geroch-Hansen mass multipole moments reduce to
the Newtonian multipole moments $I^{<i_1\cdots i_l>}$ defined in Eq.~(\ref{newtmultexpstf}), with a different
normalization: $M^{<i_1\cdots i_l>}=(2l-1)!!I^{<i_1\cdots i_l>}$ (STF mass monopoles in Thorne's
paper~\cite{Thorne:1980ru}, instead, have the same normalization as the Newtonian multipole moments). 

When extra fields are present, they correspond to new potentials and then to new sets of multipole moments.  For
instance, a scalar field $\phi$ brings a potential $\Phi_S=\phi$~\cite{Pappas:2014gca} which yields a set of scalar
multipoles $S^{<i_1\cdots i_l>}$.  It should be noted that the Ricci terms ${\cal R}_{ij}$ in the definition of
multipole moments induce a mixing of the moments associated to different fields.

When the spacetime is symmetric with respect to an axis $\hat k$, 
\begin{eqnarray}
 M^{<i_1\cdots i_l>}&=&(2l-1)!!M_lk^{<i_1}\cdots k^{i_l>}\nonumber\\
 S^{<i_1\cdots i_l>}&=&(2l-1)!!S_lk^{<i_1}\cdots k^{i_l>}\,.\label{momaxis}
\end{eqnarray}
The mass and current STF moments reduce to two sets of scalar moments $M_l$, $S_l$. The Geroch-Hansen procedure can be
cast in a simpler form~\cite{Fodor:1989aa}, exploiting the powerful Ernst potential formalism. Indeed, it can be shown
that the complex combination $\zeta=\Phi_M+{\rm i}\Phi_J$ is the (secondary) Ernst potential, and Einstein's equations
can be formulated in terms of the Ernst potential.

\subsection*{Multipole moments in general relativity: Thorne's expansion}
\addcontentsline{toc}{subsection}{A3: Multipole moments in general relativity: Thorne's expansion}

The multipole moments of a stationary, isolated object can also be defined in terms of the asymptotic behaviour of the
spacetime metric. Indeed, as shown by Thorne~\cite{Thorne:1980ru}, the (asymptotic) spacetime metric can be expanded in
inverse powers of a suitable radial coordinate; the coefficients of this expansion can be interpreted as the multipole
moments:
\begin{eqnarray}
  g_{00}&=&-1+\frac{2M}{r}+\sum_{l\ge2}\frac{1}{r^{l+1}}\left(\frac{2}{l!}M^{<a_1\cdots a_l>}
  n^{a_1}\cdots n^{a_l}+(l'<l~\hbox{harmonics})  \right)\nonumber\\
  g_{0j}&=&-2\sum_{l\ge1}\frac{1}{r^{l+1}}\left(\frac{1}{l!}\epsilon^{jka_l}S^{<ka_1\dots a_{l-1}>}n^{<a_1}\cdots n^{a_l>}
  +(l'<l~\hbox{harmonics})  \right.\nonumber\\
  &&+(l\!-\!\hbox{harmonics with parity }(-1)^l)\Bigg)\,.
  \label{Texpansion}
\end{eqnarray}
The coordinate systems in which the spacetime metric has the form~(\ref{Texpansion}) are called ACMC coordinates: they
are asymptotically cartesian and (when the source is weak-field and covered by the same coordinate system) the origin of
the coordinate system lies at the center of mass of the source. It has been shown~\cite{gursel1983multipole} that the
definitions of multipole moments by Thorne and by Geroch-Hansen are equivalent. Actually, the moments appearing
in~\cite{Thorne:1980ru} have different normalization; however, for simplicity of notation, in  Eqns.~(\ref{Texpansion}) (and
throughout this Review) 
we have used the normalizations of Geroch-Hansen, which are also those adopted in most of the recent literature on the
subject.

As discussed above, when the spacetime is symmetric with respect to an axis $\hat k$, the multipole moments
$M^{<a_1\cdots a_l>}$, $S^{<a_1\cdots a_l>}$ reduce to the scalar quantities $M_l$, $S_l$ (see
Eqns.~(\ref{momaxis})). Using the normalization properties of STF
tensors~\cite{blanchet1986radiative,poisson2014gravity}, it can be shown that
\begin{eqnarray}
 M^{<a_1\cdots a_l>}n^{<a_1}\cdots n^{a_l>}&=&l!\,M_lP_l(\cos\theta)\nonumber\\
 \epsilon^{jka_l}S^{<ka_1\dots a_{l-1}>}n^{<a_1}\cdots n^{a_l>}&=&
 (l-1)!\,\epsilon^{ijk}n^ik^k\,S_lP'_l(\cos\theta)\nonumber\\
  \end{eqnarray}
where $\cos\theta=\hat n\cdot\hat k$, and primes denote derivatives with respect to $\cos\theta$.  The metric expansion
can then be written as
\begin{eqnarray}
  g_{00}&=&-1+\frac{2M}{r}+\sum_{l\ge2}\frac{1}{r^{l+1}}\left(M_lP_l(\cos\theta)+(l'<l~\hbox{harmonics})  \right)\nonumber\\
  g_{0j}&=&-2\sum_{l\ge1}\frac{1}{r^{l+1}}\left(\epsilon^{ijk}n^ik^{k}\frac{S_l}{l}P'_l(\cos\theta)+(l'<l~\hbox{harmonics})
  \right)
      \label{Taxexpansion}
\end{eqnarray}
(note that when the spacetime is symmetric with respect to the axis $\hat k$, the vector harmonics with parity $(-1)^l$
in the expansion of $g_{0j}$ - see Eq.~(\ref{Texpansion}) - identically vanish). 
In polar coordinates $\hat k=(0,0,1)$, $\hat n=(\sin\theta\cos\phi,\sin\theta\sin\phi,\cos\theta)$ and
$\epsilon^{ijk}n^ik^kdx^j=r\sin^2\theta d\phi$, therefore Eqns.~(\ref{Taxexpansion}) yield Eqns.~(\ref{tmoments}).

If the source can be covered by de Donder coordinates (in which ${\bar
  h}^{\mu\nu}\equiv-(-g)^{1/2}g^{\mu\nu}+\eta^{\mu\nu}$ satisfies ${\bar h}^{\mu\nu}_{~~,\nu}=0$, and Einstein's
equations can be written as flat-space wave equations for ${\bar h}^{\mu\nu}$), it is possible to define an ''effective''
stress-energy tensor $\tau^{\mu\nu}$, which is the source of the wave equations.  In these coordinates, Thorne's
multipole moments can be expressed as integrals of $\tau^{\mu\nu}$ over the source~\cite{Thorne:1980ru}:
\begin{eqnarray}
M^{<a_1\cdots a_l>}&=&(2l-1)!!\int\tau^{00}x^{<a_1}\cdots x^{a_l>}d^3x\nonumber\\
S^{<a_1\cdots a_l>}&=&\frac{2l(2l-1)!!}{l+1}\int x^i\tau^{0j}\epsilon^{ij<a_l}x^{a_1}\cdots x^{a_{l-1}>}d^3x\label{intsource}\,.
\end{eqnarray}
When the spacetime is symmetric with respect to an axis $\hat k$, Eqns.~(\ref{intsource}) give
\begin{eqnarray}
  M_l&=&\int\tau^{00}r^lP_l(\cos\theta)d^3x\nonumber\\
  S_l&=&\frac{2}{l+1}\int\epsilon^{ijk}\tau^{0j}n^ik^kr^lP'_l(\cos\theta)d^3x\,.\label{intaxsource}
\end{eqnarray}
In the case of a weak-field source, $\tau^{00}=\rho$, $\tau^{0j}=\rho v^j$, and Eqns.~(\ref{intaxsource}) reduce
to~\cite{Ryan:1996nk,Stein:2013ofa}
\begin{eqnarray}
  M_l&=&\int\rho r^lP_l(\cos\theta)d^3x\nonumber\\
  S_l&=&\frac{2}{l+1}\int\rho v^\phi r^lP'_l(\cos\theta)\sin^2\theta d^3x\,.
\end{eqnarray}

\bibliographystyle{iopart-num}
\bibliography{refs}
\end{document}